\documentclass[
superscriptaddress,
reprint,
amsmath,amssymb,
aps,
prl,
]{revtex4-2}

\usepackage{graphicx}
\usepackage{dcolumn}
\usepackage{bm}
\usepackage{hyperref}
\usepackage[utf8]{inputenc}

\newcommand{\fig}[1]{Fig.~\ref{#1}}

\newcommand{\Fig}[1]{Figure~\ref{#1}}

\newcommand{\eq}[1]{Eq.~\eqref{#1}}

\newcommand{\Eq}[1]{Equation~\eqref{#1}}

\begin{document}

\title{Robust purely optical signatures of Floquet states in laser-dressed crystals}

\author{Vishal Tiwari}
\affiliation{Department of Chemistry, University of Rochester, Rochester, New York 14627, USA}

\author{Roman Korol}
\affiliation{Department of Chemistry, University of Rochester, Rochester, New York 14627, USA}

\author{Ignacio Franco}
\affiliation{Department of Chemistry, University of Rochester, Rochester, New York 14627, USA}
\affiliation{Department of Physics, University of Rochester, Rochester, New York 14627, USA}

\email{ignacio.franco@rochester.edu}

\date{\today} 

\pagebreak

\begin{abstract}

Strong light-matter interactions can create non-equilibrium materials with  on-demand novel functionalities. For periodically driven solids, the Floquet theorem provides  the natural states to characterize the physical properties of these laser-dressed systems. However,  signatures of the Floquet states are needed, as common experimental conditions, such as pulsed laser excitation and dissipative many-body dynamics, can disrupt their formation and survival. Here, we identify a tell-tale signature of Floquet states in the linear optical response of laser-dressed solids that remains prominent even in the presence of strong spectral congestion of bulk matter. To  do so, we introduce a computationally efficient strategy based on the Floquet formalism to finally capture the  full frequency-dependence in the optical response properties of realistic laser-dressed crystals,  and use it investigate the Floquet engineering in a first-principle model for ZnO of full dimensionality. The computations reveal intense, spectrally isolated, laser-controllable, absorption/stimulated emission features at mid-infrared energies present for a wide range of laser-driving conditions that arise due to the hybridization of the Floquet states. As such, these spectral features open a purely optical pathway to investigate the birth and survival of Floquet states while avoiding the experimental challenges of fully reconstructing the band structure.

\end{abstract}

\maketitle

Strong coupling of matter with light provides unprecedented opportunities for manipulating the  physical properties of materials \cite{Oka2019, Curtis2019, delaTorre2021, Budden2021, Bao2021,GarciaVidal2021, Tyulnev2024}.  In particular, time-periodic lasers create laser-dressed solids, with emerging \emph{non-equilibrium} properties that are most naturally understood through \emph{stationary} populations of Floquet-Bloch states \cite{Faisal1997, Wang2013, Shan2021, Earl2021, Aeschlimann2021, Tiwari2023}, that satisfy both Floquet and Bloch theorem.

Because experimental realizations of Floquet engineering rarely satisfy Floquet theorem --that requires unitary dynamics under strict periodic driving-- there has been a significant recent interest in identifying tell-tale signatures of the Floquet-Bloch states that can signal their birth and evolution under effects such as pulsed lasers and dissipative many-body dynamics that can disrupt their formation and survival. For this, time- and angle-resolved photoemission spectroscopy (Tr-ARPES) has emerged as a useful technique \cite{Boschini2024} to reconstruct the band structure and visualize the characteristic Floquet replicas \cite{Wang2013, Aeschlimann2021, Ito2023, Zhou2023a, Zhou2023, Choi2024, Merboldt2024}, distorted copies of the electronic states separated in energy by multiples of the driving laser photon energy. 


As an alternative, purely optical absorption experiments would be preferred as they are more common and often easier to realize experimentally. In the linear optical  response of laser-dressed matter,  the Floquet replicas lead to  absorption sidebands  that have been computationally illustrated in model one-dimensional solids \cite{Tiwari2023} and experimentally demonstrated in systems of low dimensionality\cite{Deng2015, Koski2018, Park2022, Kobayashi2023, Boos2024, Li2024}. However, in bulk materials, although the Floquet replicas have been observed in Tr-ARPES \cite{Zhou2023, Zhou2023a}, they have remained elusive in the optical response, presumably because they are obscured by spectral congestion.  This has prevented the use of  optical absorption spectroscopy  to study the emergence and evolution of Floquet  states in materials.

Systematic progress requires  developing a non-equilibrium theory of the linear response properties of laser-dressed solids and quantifying such a response in   atomistically detailed models. According to the fluctuation-dissipation theorem,  for near-equilibrium matter the response  at frequency $\omega$ is dictated by the Fourier transform of one-time $\tau=t_2-t_1$ correlation functions $C(\tau)$ of the thermal state \cite{Kubo1963, Kubo1957, Mukamel}.  By contrast, for non-equilibrium laser-dressed materials,   the  response at frequency $\omega$ is now determined by a double Fourier transform of  two-time correlation functions $C(t_1, t_2)$ because the time-translational symmetry is broken. This quantity is challenging to evaluate as it requires propagating the many-body quantum dynamics for long times forward and back for each pair of times $t_1$ and $t_2$ and for each frequency $\omega$ \cite{Gu2018, Tiwari2023}.

In this Letter, we demonstrate the first computations of the  non-equilibrium optical absorption spectra of a realistic  laser-dressed three-dimensional solid --ZnO described using a first-principle density functional theory (DFT) Hamiltonian. As detailed in our companion paper \cite{Tiwari2025}, these computations are now feasible due to the development of: (i) a Floquet-based approach to efficiently compute the full frequency dependence of the optical response  of laser-dressed solids; (ii) a truncated velocity gauge for the light-matter interactions that enables the use of first-principle Hamiltonians and overcomes well-known convergence issues of the velocity gauge with the number of bands \cite{Virk2007, Yakovlev2017, Taghizadeh2017}; (iii) the massively parallel computational implementation of the theory and; (iv) its integration with electronic structure codes for solids \cite{code}. Our computations reveal robust, laser-controllable, spectrally isolated, intense optical features at mid-infrared frequencies that provide a purely optical tell-tale signature of the emergence of Floquet states. These features can be used to monitor their birth and evolution in solids through optical absorption spectroscopy.

We simulate a drive-probe physical situation in which a continuous wave laser  of arbitrary intensity and frequency drives a solid of arbitrary band-structure and dimensionality out of equilibrium. The effective absorption properties of this laser-dressed solid are then probed by a weak continuous wave laser source. We treat the drive laser exactly through a Floquet formalism while the influence of the probe  is  captured  to first-order in perturbation theory. In our companion paper \cite{Tiwari2025}, we detail the underlying theory and computational framework  to compute the frequency-dependent optical absorption spectra of laser-dressed solids using first-principle Hamiltonians.

Briefly, the total Hamiltonian is $ \hat{H}(t) = \hat{H}_{\mathrm{LD}}(t) + \hat{H}_{\mathrm{p}}(t)$, where $\hat{H}_{\mathrm{LD}}(t)$ describes the  material and its interaction with  a drive laser $\mathbf{E}_{\mathrm{d}}(t)=\hat{\mathbf{e}}_{\mathrm{d}}E_{\mathrm{d}}\cos(\Omega t)$ with period $T=2\pi/\Omega $, amplitude $E_{\mathrm{d}}$ and polarization $\hat{\mathbf{e}}_{\mathrm{d}}$. In turn, $\hat{H}_{\mathrm{p}}(t)$ describes the interaction with the probe laser $\mathbf{E}_{\mathrm{p}}(t) = \hat{\mathbf{e}}_{\mathrm{p}} E_{\mathrm{p}} \cos(\omega t)$ of amplitude  $E_{\mathrm{p}}$, frequency $\omega$ and polarization $\hat{\mathbf{e}}_{\mathrm{p}}$. We consider the effect of the drive laser exactly while the influence of the probe  is  captured  to first-order in perturbation theory.  In second quantization, $\hat{H}_{\mathrm{LD}}(t) = \sum_{\mathbf{k} \in \mathrm{BZ}}\sum_{u,v} \langle \psi_{u\mathbf{k}}|\mathcal{\hat{H}}_{\mathrm{LD}}(t)|\psi_{v\mathbf{k}}\rangle \hat{c}^{\dagger}_{u\mathbf{k}}\hat{c}_{v\mathbf{k}} $, where $\mathcal{\hat{H}}_{\mathrm{LD}}(t)$ is the single-particle laser-dressed Hamiltonian. Here, $\hat{c}^{\dagger}_{u\mathbf{k}} (\hat{c}_{u\mathbf{k}})$  creates (annihilates) a fermion in  Bloch state $|\psi_{u \mathbf{k}} \rangle = V^{-1/2} e^{i \mathbf{k} \cdot \hat{\mathbf{r}}} | u \mathbf{k} \rangle$ with band index $u$, and crystal momentum $\mathbf{k}$ in the first Brillouin zone (BZ), where $| u \mathbf{k} \rangle $ is the Bloch mode and $V$ the crystal's volume. Since $\mathcal{\hat{H}}_{\mathrm{LD}}(t) = \mathcal{\hat{H}}_{\mathrm{LD}}(t+T) $ is time-periodic, we can invoke Floquet theorem where now the Floquet-Bloch states $|\Psi_{\alpha \mathbf{k}}(t)\rangle = V^{-1/2} e^{-iE_{\alpha  \mathbf{k}} t/ \hbar} e^{ i\mathbf{k} \cdot \hat{\mathbf{r}} } | \Phi_{\alpha \mathbf{k}} (t) \rangle $ are solutions to the time-dependent Schr\"odinger equation $(i\hbar \frac{\partial }{\partial t} |\Psi_{\alpha \mathbf{k}}(t)\rangle = \mathcal{\hat{H}}_{\mathrm{LD}}(t) |\Psi_{\alpha \mathbf{k}}(t)\rangle )$.  The quasienergies $\{E_{\alpha  \mathbf{k}}\}$  and Floquet-Bloch modes  $\{| \Phi_{\alpha \mathbf{k}} (t) \rangle \}$ are solutions to the eigenvalue problem $(i\hbar \frac{\partial }{\partial t} - \mathcal{\hat{H}}_{\mathrm{LD}}(t)) | \Phi_{\alpha \mathbf{k}} (t) \rangle = E_{\alpha  \mathbf{k}} | \Phi_{\alpha \mathbf{k}} (t) \rangle $ in Sambe space \cite{Sambe1973}. The quasienergies are uniquely defined in the first Floquet-Brillouin zone (FBZ) such that $ -\hbar\Omega/2 \leq E_{\alpha  \mathbf{k}} < \hbar\Omega/2 $ for $\mathbf{k} \in $ BZ. The Floquet-Bloch modes are time-periodic with period $T$ and space-periodic with period $\mathbf{R}$ --the lattice vector of the crystal.

The non-equilibrium optical absorption coefficient of a laser-dressed crystal at probe photon energy $\hbar\omega$ is \cite{Tiwari2025}:
\begin{align}
\nonumber
    & A(\omega)     =  \frac{e^{2}\pi}{ m_{e}^2 \epsilon_{0}c n_{r} V \omega}\sum_{\mathbf{k} \in \mathrm{BZ} }\sum_{\alpha, \beta}\sum_{n}\Lambda_{\alpha \beta \mathbf{k}}|\mathcal{Z}_{\alpha\beta \mathbf{k}}^{(n)}|^2  \\
        \label{final}
 &  \times   [\delta(E_{\alpha\beta \mathbf{k}}+n\hbar\Omega-\hbar \omega)-\delta(E_{\alpha\beta \mathbf{k}}+n\hbar\Omega+\hbar \omega)] .
\end{align}
Here,   $\epsilon_{0}$ is vacuum's electric permittivity, $c$ the speed of light and $n_{r}$ the material's refractive index. The quantity $ E_{\alpha\beta \mathbf{k}} = E_{\alpha \mathbf{k}}-E_{\beta \mathbf{k}} $ is the Bohr transition energy between Floquet-Bloch mode $\alpha$ and $\beta$ at  crystal momentum $\mathbf{k}$. The population factor
$
\Lambda_{ \alpha \beta \mathbf{k} } = V^{-4}\sum_{u', u}| \langle u\mathbf{k}|\Phi_{\beta \mathbf{k}}(t_{0})\rangle |^2 |\langle\Phi_{\alpha \mathbf{k}}(t_{0})|u'\mathbf{k}\rangle |^2 
  \bar{n}_{u \mathbf{k}}(1-\bar{n}_{u' \mathbf{k}}) ,$
where  $\bar{n}_{u \mathbf{k}} = \langle \psi (t_{0}) | \hat{c}^{\dagger}_{u\mathbf{k}} \hat{c}_{u\mathbf{k}} | \psi (t_{0}) \rangle $ is the occupation number of band $u$ and crystal momentum $\mathbf{k}$ of many-body state $|\psi (t_{0}) \rangle$ at preparation time $t_{0}$. In turn, $\mathcal{Z}_{\alpha\beta \mathbf{k}}^{(n)}$ is the $n$-th Fourier component of the  truncated momentum matrix element $ \mathcal{Z}_{\alpha\beta \mathbf{k}}(t)=V^{-1} \langle\Phi_{\alpha \mathbf{k}}(t)| e^{-i \mathbf{k} \cdot \hat{\mathbf{r}}} \hat{z}(t) e^{i \mathbf{k} \cdot \hat{\mathbf{r}}} | \Phi_{\beta \mathbf{k}}(t)\rangle 
$ between the Floquet-Bloch mode $\alpha$ and $\beta$ at crystal momentum $\mathbf{k}$, where $ \hat{z}(t)$ is the truncated momentum operator (cf. Eq. 14 in Ref. \cite{Tiwari2025}).

\begin{figure*}[htbp!]
    \centering
    \includegraphics[width=0.98\textwidth]{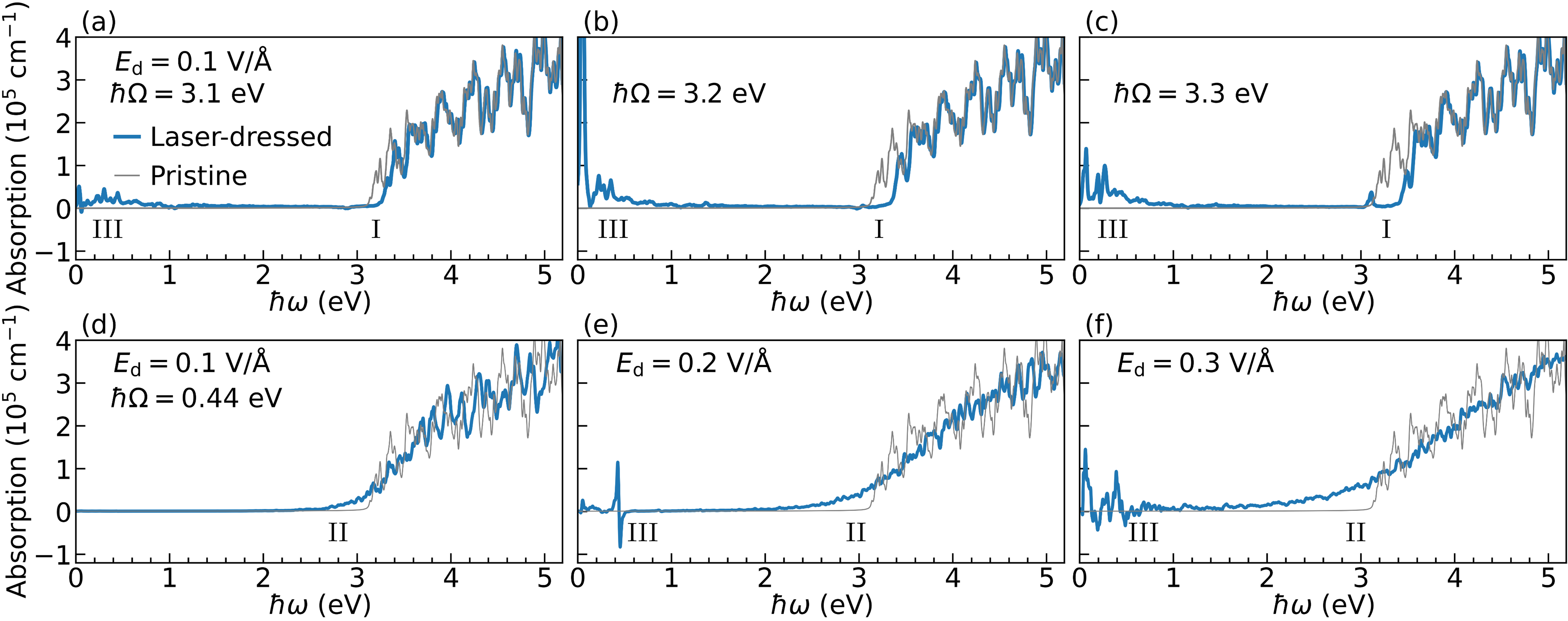}
    \caption{Frequency ($\hbar\omega$) dependence of the optical absorption spectrum $A(\omega) n_{r}$ for  ZnO dressed with near-resonant (a)-(c) and off-resonance (d)-(f) light  of photon energy $\hbar\Omega$ and amplitude $E_{\mathrm{d}}$ (blue lines). Gray lines: spectra of pristine solid. Floquet engineering leads to strong modification of the optical response, including band-gap renormalization (feature I), below band-gap absorption (II) and  the emergence of mid-IR features (III). }
    \label{opticalabs}
\end{figure*}

\Eq{final} is valid for any arbitrary  time-periodic driving laser and for  any space-periodic material including semiconductors, semi-metals, insulators and quantum materials. When interfaced with DFT computations, it captures electronic correlations at preparation time.
Remarkably, its structure  is analogous to the near-equilibrium optical response \cite{Dresselhaus2018} but with the Floquet-Bloch modes playing the role of the pristine eigenstates. Specifically, optical absorption in laser-dressed solids arise due to transitions among Floquet-Bloch modes  $\beta \rightarrow \alpha$ separated by $n$ Floquet-Brillouin zones with Bohr transition energy $E_{\alpha \beta \mathbf{k}} + n\hbar \Omega$. Here $n= 0, \pm 1, \cdots, \pm n_{\mathrm{F}}$ where $2n_{\mathrm{F}} +1 $ is the maximum number of FBZs  needed for convergence. The first term  represents absorption, the second  stimulated emission. Transition intensities  are determined by the population factor $\Lambda_{\alpha \beta \mathbf{k}}$ (that guarantees transitions occur from an occupied  to an empty state) and the momentum matrix element $\mathcal{Z}_{\alpha\beta \mathbf{k}}^{(n)}$ (that are dependent on the number $n$ of FBZs separating the modes). Interestingly, in this non-equilibrium context, transition matrix elements and population factors depend on the strength and nature of the driving laser (instead of being properties inherent to the material only), and optical transitions occur among states that can be separated by a variable number of FBZs leading to absorption sidebands.


To find optical signatures of Floquet states, we use  \eq{final} to compute the absorption spectrum for  laser-dressed wurtzite ZnO crystal using  a first-principle Hamiltonian and  the computational strategy detailed in Ref. \cite{Tiwari2025}. The large $\sim 3.2$ eV  ZnO  band-gap enables resonant and non-resonant laser-dressing using experimentally accessible frequencies \cite{Ghimire2010, Ghimire2011}.  The DFT electronic structure of ZnO was computed using Quantum Espresso \cite{Giannozzi2017} with Hubbard (DFT+U) corrections needed to reproduce  experimental band gaps \cite{Janotti2006, Calzolari2011, Spencer2022} followed by a Wannier interpolation using Wannier90 \cite{Pizzi2020} that enables efficient computation of \eq{final}.  The employed 22-Wannier model accurately reproduces the DFT band structure in a 25 eV energy range (see Fig. S1 in Supplemental Material (SM)\cite{supmat}) and, together with $n_{\mathrm{F}}= 301$ and a dense $\mathbf{k}$-space grid grid of $(30)^3$, provides converged results for the non-equilibrium optical absorption. Throughout, since electron-phonon interactions introduce Franck-Condon like progressions \cite{Szidarovszky2019}, in our computations spectral features are Lorentzian broadened with width 0.02 eV.  Further, transitions with energy less than 0.06 eV are not reported to account for the limits of the finite BZ discretization. We take  the drive and probe to be linearly polarized along the crystal's $c$-axis. Other polarization directions yield qualitatively similar physics.  Additional computational details are included in the SM \cite{supmat}.

Figure \ref{opticalabs} shows the optical absorption spectra of pristine ZnO (gray lines) and  laser-dressed (blue lines) ZnO with field amplitudes $E_{\mathrm{d}}=0.1-0.3$ V/\r{A} (intensities 0.1-1.2 TW/cm$^2$) in the intermediate regime of light-matter interaction (non-ionizing, but non-perturbative) and with frequency chosen to be near-resonance with the band-edge ($\hbar\Omega = 3.1-3.3$ eV, Fig.\ref{opticalabs}a-c) and off-resonance ($\hbar\Omega = 0.44$ eV, Fig.\ref{opticalabs}d-f). The pristine spectrum shows a sharp band-edge at $ 3.2$ eV  as observed experimentally \cite{Ghimire2011, Spencer2022}. Floquet engineering leads to strong modification of the optical response.

In the near-resonance case (\fig{opticalabs}a-c), the laser-dressing strongly suppresses the absorption for $\hbar\omega \approx \hbar\Omega$ and leaves the spectrum for $\hbar\omega>4$ eV unchanged (feature I). This suppression is due to band-gap renormalization created by  hybridization of the Floquet-Bloch states (see associated Tr-ARPES evidence  in Ref. \cite{Zhou2023}). More explicitly, when the solid is  resonantly driven, the first  Floquet replica of the valence (conduction) band overlaps  with the conduction (valence) band. This leads to band  hybridization that opens energy gap at $\hbar\omega \approx \hbar\Omega$, thus removing transitions at this frequency. This band-gap renormalization  is an optical signal of the formation of Floquet-Bloch states as it occurs due to their hybridization. However, the challenge with isolating this feature in the optical spectra is that it emerges for $\hbar\omega$ around the band-edge  where it can be obscured by excitonic features \cite{Reynolds1999, zgr2005, Rohlfing2000} which are not included in the model.


For off-resonance driving (\fig{opticalabs}d-f), the  laser-dressing leads to below band-edge absorption features $\hbar\omega < 3.2$ eV (II) that spread to even lower energies as the field strength is increased.  This so-called dynamical Franz-Keldysh Effect \cite{Jauho1996} can be attributed to transitions occurring among the Floquet replicas of valence and conduction bands \cite{Tiwari2023}. For off-resonance drive laser, multiple Floquet  replicas are formed leading to transitions with  energies lower than the field-free band-edge. Further, since the replicas are separated from each other by the drive photon energy, they can lead to absorption sidebands. Both the below-band gap features and the  sidebands are optical signatures of  Floquet  states. However, as the band renormalization,  they are also challenging to observe. Below-band gap absorption overlaps in energy with dominant exciton features. In turn, while the absorption sidebands are expected from \eq{final}, and clearly visible in systems of low-dimensionality \cite{Kobayashi2023} including one-dimensional model crystals \cite{Tiwari2023}, they  are obscured in \fig{opticalabs}d-f by the  spectral congestion of the three-dimensional material.

\emph{Are there any features in the laser-dressed optical spectrum that remain visible in realistic systems?}

Interestingly, \fig{opticalabs} shows prominent absorption and stimulated emission features (III) at mid-infrared frequencies ($\hbar\omega < 0.6$ eV) that emerge under both near-resonance (\fig{opticalabs}a-c) and off-resonance (\fig{opticalabs}e-f) driving. These features are well separated from the band-edge or vibrational features, and are thus not obscured by possible excitonic contribution or  spectral congestion. Further, their emergence is robust to changes in the drive laser frequency and amplitude.

To clarify their physical origin, we developed a minimal one-dimensional two-band model that captures this phenomenon  from a projection of the  highest valence and the lowest conduction band of ZnO  constructed using  $\mathbf{k \cdot \mathbf{p}}$ perturbation theory \cite{Haug2009} (see SM\cite{supmat} for model parameters). This model system  also shows low-frequency transitions for the same drive laser parameters as ZnO (see Fig. S3 in the SM\cite{supmat}).  However, its reduced dimensionality enables to more clearly identify the essential physics.

\begin{figure}[htbp!]
    \centering
    \includegraphics[width=0.48\textwidth]{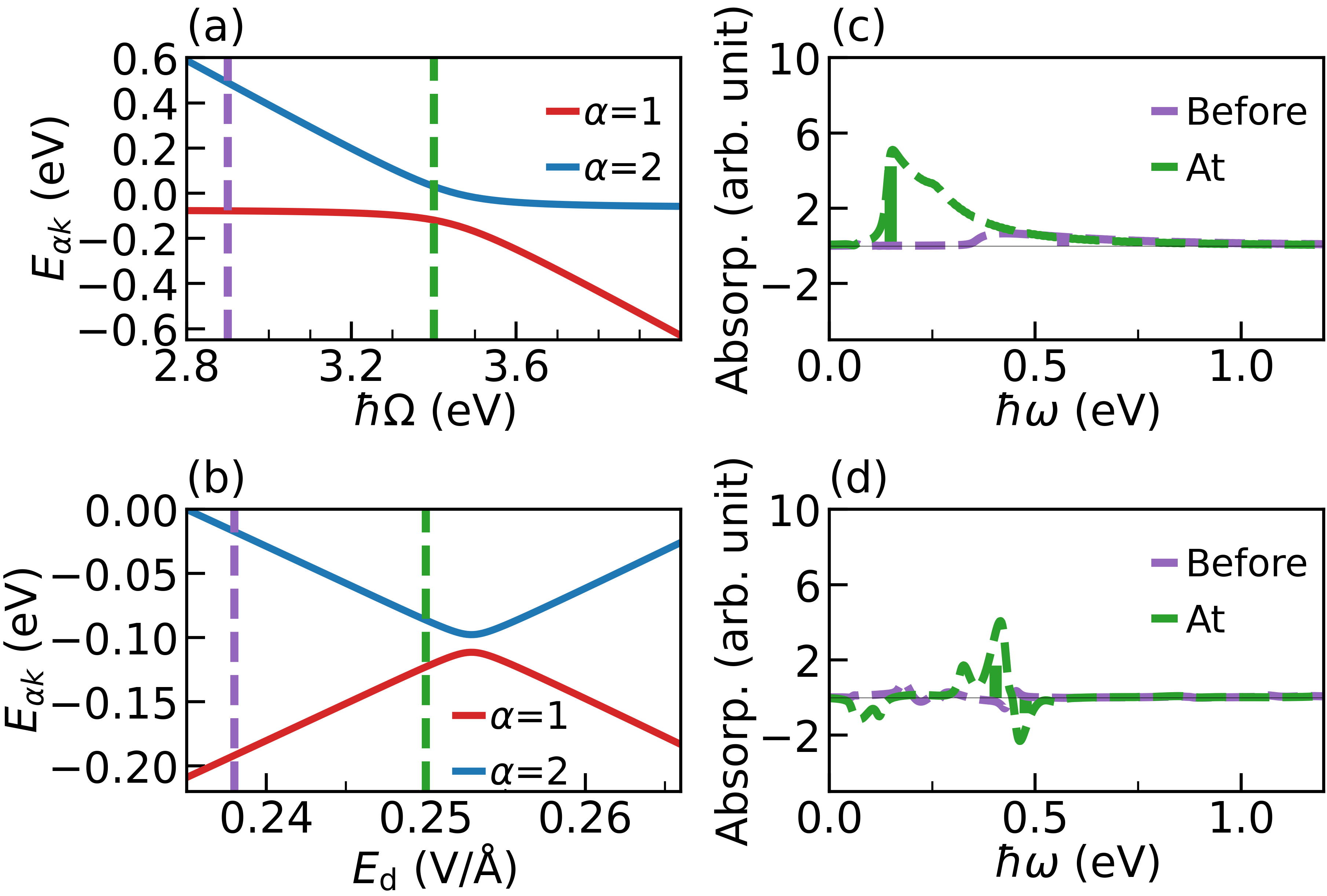}
    \caption{Low-frequency transitions in the optical response as demonstrated in a minimal two-band model for near-resonance (top row) and off-resonance (bottom row) driving. 
    (a-b) Quasienergies at a representative $k=0.1$ showing the avoided crossing as a function of (a) driving laser frequency $\hbar\Omega$ (with fixed $E_{\mathrm{d}} = 0.1$ V/\r{A}) and (b) amplitude $E_{\mathrm{d}}$ (with fixed $\hbar\Omega=0.44$ eV).  Vertical purple and green lines mark parameters before or at the avoided crossing, respectively. Panels (c) [or (d)] show the absorption spectra corresponding to (a) [or (b)] before (purple) and at (green) the avoided crossing. Transition lines signal the $k=0.1$ contribution only. Note that the low-frequency transitions emerge in the avoided crossing region where Floquet states hybridize. }
    \label{resonant}
\end{figure}

As we now show, these low-frequency features arise due to formation and subsequent hybridization of the Floquet-Bloch modes. \Fig{resonant} top row shows this for drive laser frequency near-resonant to the band-edge. We plot the quasienergies of the two-band model within a FBZ for a representative $k$-point ($k=0.1$ units of $2\pi/a$, with $a=5.2$ \r{A} the unit cell length) in \fig{resonant}a  as a function of $\hbar\Omega$. The two quasienergies form an avoided crossing around $\hbar\Omega=3.4$ eV which is a signature of level hybridization. Absorption spectrum of this two-band model with driving frequency chosen before the avoided crossing  ($\hbar\Omega=2.9$ eV purple vertical line in \fig{resonant}a), shown in \fig{resonant}c (purple line) shows  no significant low-frequency feature. But when the model is driven by laser parameters chosen at the avoided crossing ($\hbar\Omega=3.4$ eV green vertical line in \fig{resonant}a), the spectrum (\fig{resonant}c green line) exhibits an intense low-frequency absorption feature centered at $\hbar\omega =0.2$ eV with strong contribution from the transition at $k=0.1$.

Similar behavior is also seen for off-resonance drive laser with $\hbar\Omega=0.44$ eV. The two quasienergies in a FBZ for  the same representative $k=0.1$ as a function of $E_{\mathrm{d}}$ in \fig{resonant}b show an avoided crossing centered around $E_{\mathrm{d}}=0.253$ V/\r{A}. When the model is driven by laser with $E_{\mathrm{d}}=0.238$ V/\r{A} before the avoided crossing (purple vertical line in \fig{resonant}b), the spectrum shown in \fig{resonant}d (purple line) shows no significant low-frequency features in general and faint transition for $k=0.1$ in particular.  But when the  system is driven by  a  laser with $E_{\mathrm{d}}=0.25$ V/\r{A} at the avoided crossing (green vertical line in \fig{resonant}b), the spectrum shown in \fig{resonant}d (green line) shows intense absorption and stimulated emission features. This shows that in both resonantly and non-resonantly driven solid, the hybridization of the Floquet-Bloch modes indicated by the opening of avoided crossings induces low-frequency transition signals.

\begin{figure}[htbp!]
    \centering
    \includegraphics[width=0.48\textwidth]{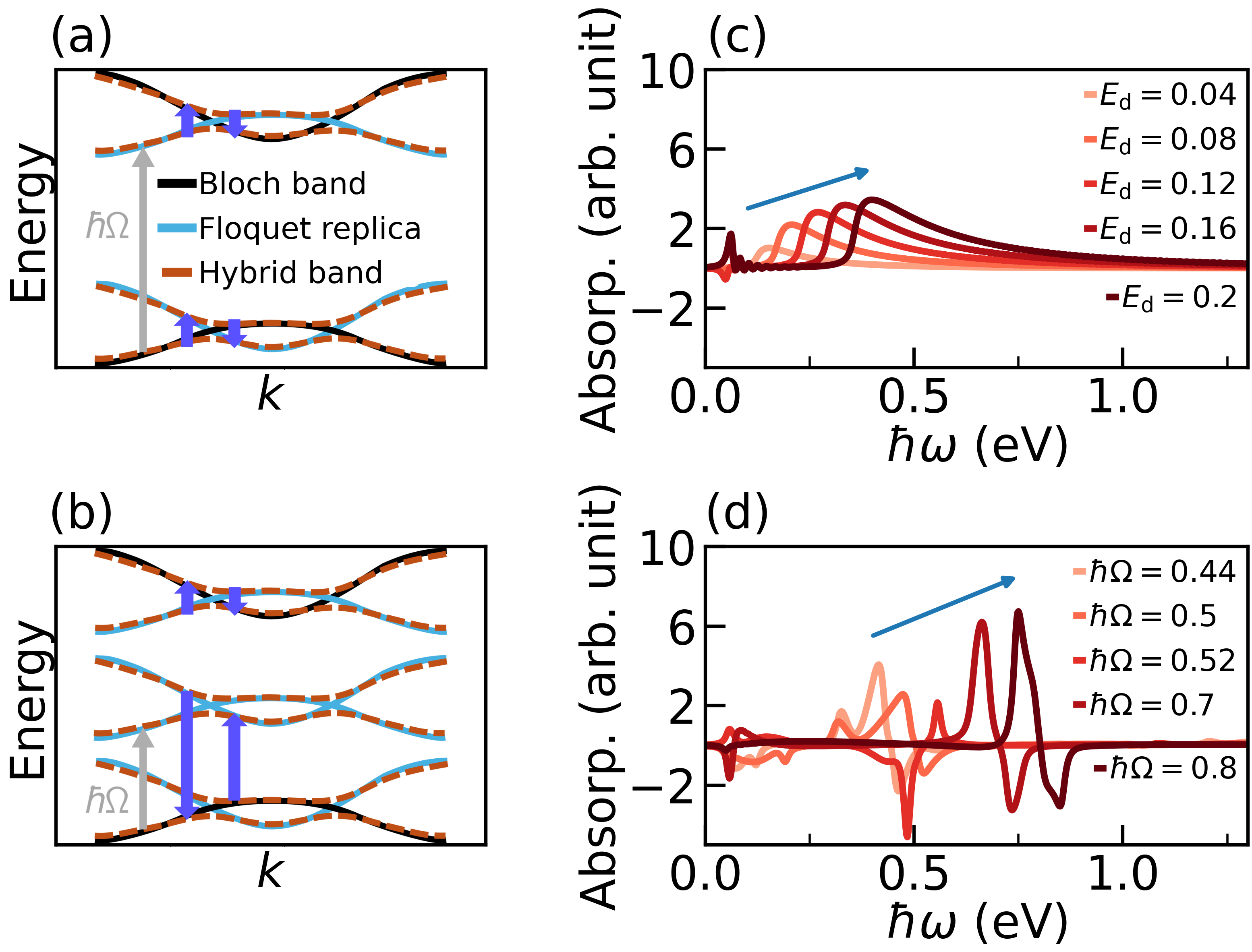}
    \caption{(a)-(b) Schematic of the laser-dressed band structure showing the emergence of the  low-frequency transitions due to Floquet hybridization for (a) resonant  and (b) non-resonant driving. Purple arrows signal possible low-frequency transitions. These transitions blue-shift with increasing driving-laser's (c)  amplitude $E_{\mathrm{d}}$ in V/\r{A} for resonant-driving ($\hbar\Omega=3.1$ eV) and (d) frequency $\hbar \Omega$ in eV ($E_{\mathrm{d}}=0.25$ V/\r{A}) for non-resonant driving.}
    \label{diagram}
\end{figure}

\Fig{diagram}a-b summarize the emergent physical picture. Laser-dressing leads to the formation of  Floquet replicas (blue lines) of the valence and conduction band (black lines) that can energetically overlap either with the original Bloch bands or their replicas. This induces hybridization of the Floquet-Bloch states and formation of hybrid bands (red lines) that display avoided crossings as a function of laser parameters (\fig{resonant}a-b).  Probing this laser-dressed band-structure leads to  intense low-frequency transitions as indicated by the purple arrows. For the resonant case \fig{diagram}a, the low-frequency features arise because of intra FBZ ($n=0$  in \eq{final}) optical transitions. In turn, for the non-resonant case \fig{diagram}b, such features can arise because of intra ($n=0$) and inter FBZ ($n=\pm 1$) optical transitions. For the inter FBZ transition $n=1$ leads to absorption and $n=-1$ leads to stimulated emission resulting in the characteristic double peak  in \fig{resonant}d around $\hbar\omega \approx \hbar\Omega$ (see also \fig{opticalabs}e), with the lower frequency peak yielding absorption and the higher-frequency one stimulated emission. Higher order transitions are possible but less prominent in the spectra for this laser-dressing. All these features emerge because of the hybridization of Floquet-Bloch modes, see \fig{resonant}.

\Fig{diagram}a-b suggests that these low-frequency features are controllable simply by varying the parameters of the driving laser. This is, in fact, clearly shown in computations in the two-band model shown in \fig{diagram}c-d. 
For resonant driving (\fig{diagram}c), increasing the laser amplitude blue-shifts the low-frequency features in the spectra (blue arrow). This is because the gap due to Floquet hybridization increases with the strength of the light-matter interactions, akin to the Autler-Townes effect \cite{Autler1955, Kim2020}. In turn, for non-resonant driving (\fig{diagram}d), the low-frequency feature due to inter FBZ transition ($n=\pm 1$) blue-shift upon increasing the drive laser frequency. This controllability can be used to characterize the emergent low-frequency features, and enable their resolution with ultrafast probes.

This origin of the low-frequency features in the two-band model can be naturally extended to the realistic laser-dressed ZnO. \Fig{resonant} shows that for a single $k$-point, the two Floquet-Bloch modes form an avoided crossing in the band structure due to hybridization for certain drive laser parameters. In the realistic computations, similar hybridization occurs but for a large number of $\mathbf{k}$-vectors given that a sufficient driving laser amplitude is applied. This leads to a plethora of low-frequency transitions, some that lead to absorption while others lead to stimulated emission  depending on whether the laser parameters are before or after each specific avoided crossing. These contributions do not exactly cancel one another leading to the robust low-frequency features observed in \fig{resonant}.

In nature, these low frequency features will arise from the hybridization of  many-body Floquet replicas. For this reason, as explicitly demonstrated for a Hubbard-Holstein model \cite{Schler2018, johnston_systematic_2010,esslinger_fermi-hubbard_2010, fehske_quantum_2004,johnston_systematic_2010} in Figs. S4-S7 in the SM~\cite{supmat} using exact diagonalization techniques \cite{Hu2018, Light2007, Jordan1928, Colbert1992}, electron correlations and electron-phonon couplings influence the frequency and magnitude of these transitions as they modulate the many-body Floquet states but do not obscure the effect.  Further, the effect is robust to spectral broadening.  In fact, as shown in Fig. S8 in the SM\cite{supmat}, increasing spectral broadening  in a wide range of energies (0.02-0.2 eV) reduces but does not obscure the main low-frequency features.

These low-frequency features are  a clear indication of the formation of Floquet states in laser-dressed materials.  Our computations suggest that to observe them in semiconductors requires moderate field intensities ($ \sim 0.1$ TW/cm$^{2}$) for lasers with  frequency near-resonant to the band-edge or stronger lasers ($\sim 2$ TW/cm$^{2}$) if dressing off-resonance. 

Dissipation, pulsed excitation and other effects that limit the applicability of the Floquet theory are expected to reduce the visibility of these low frequency features. Thus, the onset and disappearance of these laser-controllable low-frequency features can be used to test the applicability of Floquet engineering in crystals under experimentally relevant conditions. In particular, we expect that the optical controllability of these low-frequency features can be employed to characterize timescales of effects that break the Hamiltonian's time-periodicity.

To summarize, we presented the first computational demonstration of the non-equilibrium optical absorption spectrum of a laser-dressed solid (ZnO) described using first-principle  electronic structure of full dimensionality. We show that the Floquet engineering can dramatically change the optical response of a crystal. In particular,  we find  prominent absorption/stimulated emission features at mid-infrared frequencies that are robust to  drive laser parameters, survive the spectral congestion and  broadening in solids, and are well isolated from the field-free band-edge and possible excitonic transitions.   Using a minimalistic two-band model, we demonstrated that these features arise due to the hybridization of the Floquet states.  We propose that these features provide a robust, laser-controllable, purely optical tell-tale signature of the formation of Floquet states in non-equilibrium bulk matter that can be used to study the validity of Floquet engineering with varying experimental conditions (i.e. temperature, material, pulse width, etc.). As such, these spectral features open a purely optical pathway to investigate the birth and
survival of Floquet states while avoiding the experimental challenges of fully reconstructing
the band structure.

\begin{acknowledgments}
This research is based upon work supported by the National Science Foundation under Grant No. CHE-2416048. VT  acknowledges the DeRight fellowship support from the University of Rochester.
\end{acknowledgments}

\end{document}


\title{Supplementary Material: Robust purely optical signatures of Floquet  states in laser-dressed crystals}

\author{Vishal Tiwari}
\affiliation{Department of Chemistry, University of Rochester, Rochester, New York 14627, USA}

\author{Roman Korol}
\affiliation{Department of Chemistry, University of Rochester, Rochester, New York 14627, USA}

\author{Ignacio Franco}
\affiliation{Department of Chemistry, University of Rochester, Rochester, New York 14627, USA}
\affiliation{Department of Physics, University of Rochester, Rochester, New York 14627, USA}

\email{ignacio.franco@rochester.edu}

\date{\today}

\maketitle

\tableofcontents

\section{Details of the electronic structure and Non-Equilibrium Optical Absorption computation for Z\lowercase{n}O}
We obtain the electronic structure of ZnO using plane waves based density functional theory (DFT) calculations with Hubbard potential corrections (DFT+U) through Quantum ESPRESSO \cite{Giannozzi2017}. The Hubbard potential corrections are needed to obtain experimental band gaps \cite{Janotti2006, Calzolari2011, Spencer2022}.  We ignore effects due to spin-orbit coupling in ZnO. For DFT, we use the local density approximation (LDA) exchange-correlation functional with the ultra-soft pseudopotential. We first perform atomic structure relaxation and then density relaxation  on a $5\times 5\times 5$ Monkhorst-Pack $\mathbf{k}$-vector grid using wavefunction cutoff 50 Ry and charge density cutoff 400 Ry. Following the methodology detailed in Ref. \cite{Calzolari2011}, we choose Hubbard potential U values to reproduce the experimental band gaps from Ref. \cite{Ghimire2011}. Specifically we use U value of 10.0 eV on 3d orbital of Zn and 7.5 eV on 2p orbital of O in both structure and density relaxation calculations. We obtain the hexagonal lattice system with lattice vector $\mathbf{a}_{1} = a(1,0,0), \mathbf{a}_{2} = a(-1/2,\sqrt{3}/2,0), \mathbf{a}_{3} = a(0,0,c/a)$, with $a=3.25$ \r{A} and $c=5.20$ \r{A} and crystal unit cell volume $47.63$ \r{A}$^{3}$. The resulting band structure shows a band gap of 3.05 eV at $\Gamma$ point ($\mathbf{k}=(0,0,0)$).

To capture the light-matter interaction with first-principle material Hamiltonian efficiently, we employ the truncated velocity gauge \cite{Passos2018, Tiwari2025} instead of the usual velocity gauge. In this case, the light-matter interaction operator $
   ( \frac{e}{i\hbar}  )  [ \mathbf{A}(t) \cdot \hat{\bm{r}}, \hat{H}_{0}  ]+ \frac{1}{2!}  (\frac{e}{i\hbar}  )^2   \mathbf{A}(t) \cdot \hat{\bm{r}} ,  [  \mathbf{A}(t) \cdot \hat{\bm{r}} , \hat{H}_{0}  ]  ] + \cdots 
$
is now expressed as a series of nested commutators to a given order instead of it being
$e\hat{\mathbf{P}}\cdot \mathbf{A}(t)/m_{e} + e^2\mathbf{A}^2(t)/2m_{e} $. Here, $\hat{\bm{r}} = \sum_{j=1}^{M}\hat{\mathbf{r}}_{j}$ is the position and  $\hat{H}_{0}$ the Hamiltonian operator of the $M$-electron system, $\hat{\mathbf{P}}$ its conjugate momentum operator, $\mathbf{A}(t)$ the vector potential of light, $-e$ the electron's charge and $m_{e}$ its mass.  As detailed in Ref. \cite{Tiwari2025}, like the usual velocity gauge, this structure has the advantage of respecting the space-periodicity in solids  as needed to invoke Bloch theorem; yet it  overcomes its poor  convergence properties \cite{Virk2007, Yakovlev2017, Taghizadeh2017} with the number of bands and  its incompatibility with DFT Hamiltonians that employ non-local pseudopotentials.  Overall, the truncated velocity gauge  enables  tractable converged computations  of Floquet engineering in realistic solids. 

Since the computation in the truncated velocity gauge requires matrix elements of the nested commutators where the individual terms contain the position operator, we consider the solid Hamiltonian in  the generalized tight-binding model form  constructed from  maximally-localized Wannier functions (MLWFs). They allow us to evaluate the nested commutators up to all orders as, in this highly localized basis, the position operator is well-defined and easy to calculate. We construct such a generalized tight-binding model for ZnO by performing the Wannier interpolation of the DFT+U electronic structure using Wannier90 \cite{Pizzi2020}. We perform this interpolation on the $5\times 5\times 5$  $\mathbf{k}$-vector grid sampling in Brillouin zone (BZ) and obtain the material Hamiltonian and position operator matrix elements among the MLWFs.

\begin{figure}[htbp!]
    \centering
    \includegraphics[width=0.9\textwidth]{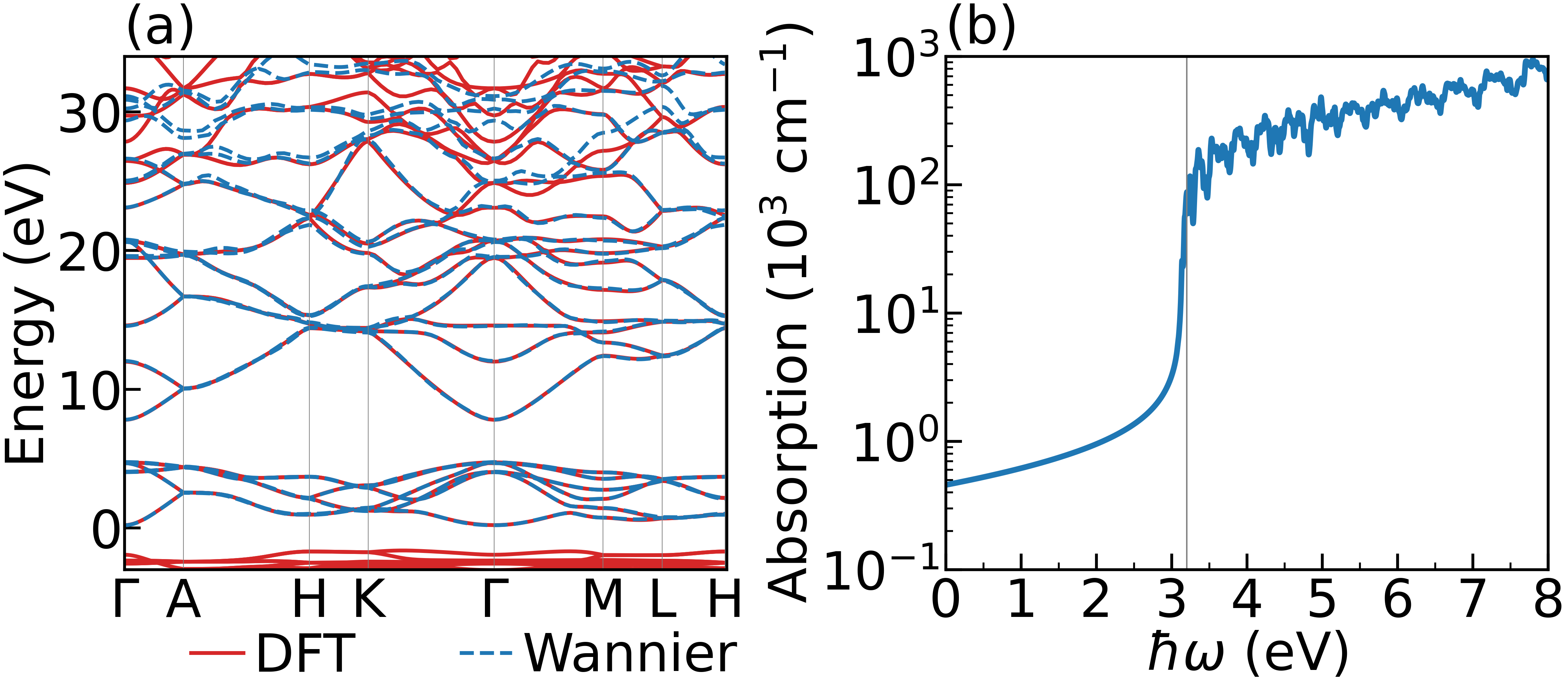}
    \caption{a) Band structure of ZnO computed using the 22-Wannier band model (blue) and the DFT+U (red). (b) Field-free absorption spectrum of ZnO using this model. The vertical line at 3.2 eV denotes the band-edge.}
    \label{fieldfree}
\end{figure}

\Fig{fieldfree}a compares the band structure along the high-symmetry points obtained using  DFT+U and the Wannier Hamiltonian constructed using 22 Wannier functions. As shown, the 22-Wannier model accurately reproduces the DFT+U band structure in a wide energy window. \Fig{fieldfree}b shows  the computed field-free optical absorption spectrum of ZnO with the BZ sampled using a  dense $(30)^3$ $\mathbf{k}$-space grid. The spectrum shows a sharp band-edge at $ 3.2$ eV (vertical black line) as observed experimentally \cite{Ghimire2011, Spencer2022}.  We find that the field-free absorption spectrum computed using Hamiltonian and position matrix elements constructed from 22 Wannier functions shows a converged band-edge as compared to a 26 Wannier function based Hamiltonian.

The computation of the optical absorption spectra of laser-dressed solids require as input the $\mathbf{k}$-vector sampling the BZ, Hamiltonian  parameters describing the realistic material in the basis of MLWFs, and the drive laser parameters ($\epsilon_{\mathrm{d}}$, $E_{\mathrm{d}}$, and $\hbar\Omega$). The $\mathbf{k}$-vectors can be obtained by imposing the Born-Von-Karman boundary condition using the lattice vectors of the material while the tight-binding parameters, for a given number of Wannier functions taken in a unit cell, are obtained from  \textsc{Wannier90} \cite{Pizzi2020}. Our code \textsc{FloqticS} \cite{code} first computes the band structure, the Bloch states, and the matrix elements of the specified number of nested commutators  in the truncated velocity gauge for the given probe and drive laser polarization ($\hat{\mathbf{e}}_{\textrm{p}}, \hat{\mathbf{e}}_{\textrm{d}}$) and for each $\mathbf{k}$-vector in the BZ. The code then constructs and diagonalizes the Floquet-Bloch Hamiltonian to obtain the quasienergies $E_{\alpha \mathbf{k}}$ and the Floquet-Bloch modes for the given drive laser amplitude and photon energy. Computations should be checked for convergence on the number of nested commutator terms, the Floquet Fourier basis states ($n_{F}$), and the number of Wannier functions defining the active computational space.

The code proceeds to calculate the Fourier components of the truncated momentum matrix elements $\mathcal{Z}_{\alpha\beta \mathbf{k}}^{(n)}$, and the population factor $\Lambda_{ \alpha \beta \mathbf{k} }$ using $\bar{n}_{u\mathbf{k}}$ as inputs (see main text). The code outputs the optical absorption spectrum of the laser-dressed material with each absorption line broadened using a Lorentzian function of a given width.

\textsc{FloqticS} parallelizes the diagonalization of the Floquet-Bloch Hamiltonian, and the calculation of the Fourier components of the truncated momentum matrix elements by distributing computations across different processors. The truncated velocity gauge enables efficient convergence of the computations with the number of bands. Both of these developments now enable to computation of non-equilibrium optical absorption properties with a fine Brillouin zone sampling for realistic solids in tractable computational time. For the computations of the optical absorption spectra of laser-dressed ZnO (Fig 1) we employ the 22-Wannier model,  23 nested commutator terms, and a basis of time-periodic functions with $n_F=301$ which provide converged computations. Each $k$-point computation takes 1.7 CPU mins (Intel Xeon Gold 6330), and overall one such spectrum shown in Fig. 1  takes 31 CPU days for each set of driving laser-parameters.

\section{2-band model using the \textbf{k.P} theory }
To develop a minimal model to capture the electronic structure of ZnO requires determining the band energies and momentum matrix elements for $k$-points sampling the BZ. For the two-band model, consider the single-particle Hamiltonian $\hat{h}_{0} = \hat{p}^2/2m_e + V(\hat{r})$, where $\hat{r}$ is the position operator, $\hat{p}$ is the conjugate momentum, and $V(\hat{r})$ is the space-periodic lattice potential. As the Hamiltonian is space-periodic, the Bloch theorem is applicable and the solution to the eigenvalue problem
\begin{equation}
\label{twobandsch}
    \hat{h}_{0} | \psi \rangle = \epsilon | \psi \rangle
\end{equation}
are $| \psi \rangle \equiv |\psi_{uk} \rangle =  e^{ik \cdot \hat{r}} | u k\rangle $, where $|u k \rangle$ is the space-periodic Bloch function. Using the  \textbf{k.P} perturbation theory \cite{Haug2009}, the Bloch function at any $k$-vector can be expanded in terms of the Bloch function at the  $\Gamma$ point such that $| u k \rangle =  \sum_{n} C_{n k}^{(u)} |n 0 \rangle$, where $|n 0\rangle$ is the Bloch function for band $n$ at $\Gamma$ point. Substituting this Bloch function definition in \eq{twobandsch} and using Bloch theorem gives
\begin{equation}
\label{kpsch}
  \sum_{n} ( \hat{h}_{k} + \frac{\hbar k\cdot \hat{p}}{m_{e}} + \frac{\hbar^2 k^2}{2m_{e}} )  C_{n k}^{(u)} |n 0 \rangle = \epsilon_{u k} \sum_{n} C_{n k}^{(u)}| n 0 \rangle ,
\end{equation}
where $h_{k} =  e^{-i k \cdot \hat{r}} \hat{h}_{0} e^{i k \cdot \hat{r}}$. Left multiplying by $|m 0 \rangle$ yields
\begin{equation}
\label{eigval}
   \sum_{n} ((\epsilon_{n0} + \frac{\hbar^2 k^2}{2m_{e}}) \delta_{n,m } + \frac{\hbar k}{m_{e}}\cdot \langle m 0 |\hat{p}| n 0 \rangle ) C_{n k}^{(u)} = \epsilon_{uk} C_{mk}^{(u)} .
\end{equation}
Once the {$\langle m 0 |\hat{p}| n 0 \rangle$} and {$\epsilon_{n0}$} values are known, \eq{eigval} can be solved to obtain the band energies $\epsilon_{uk}$ and the momentum matrix element $ \langle \psi_{uk} | \hat{p} | \psi_{vk} \rangle $.

\begin{figure}[htbp!]
    \centering
    \includegraphics[width=0.85\textwidth]{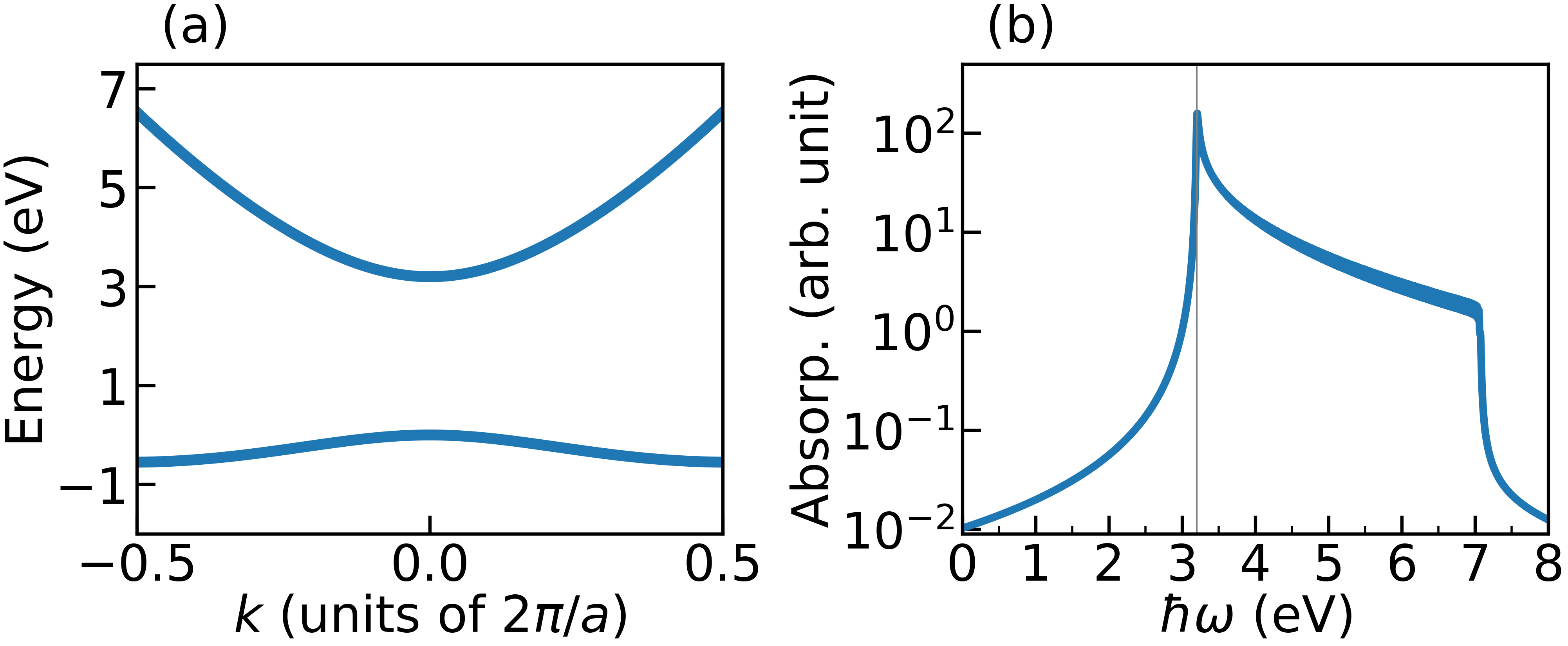}
    \caption{(a) Band structure of the two-band model and (b) its field-free absorption spectrum with model parameters obtained  using the \textbf{k.P} perturbation theory. }
    \label{kpbandstructure}
\end{figure}

To capture the ZnO electronic structure using the 2-band model, we use the parameters that reproduce the  highest valence band and the lowest conduction band of ZnO in the reciprocal space using along the $\mathbf{a}_{3}$ lattice vector (c-axis of the crystal). We take a p-like orbital ($| n 0 \rangle$ = $|p 0 \rangle$) for the valence band representing the 2p-O orbital contribution and the s-like orbital ($| m 0 \rangle = |s 0 \rangle$) for the conduction band representing the 4s-Zn orbital contribution. We take the energy $\epsilon_{p0} = 0.0$ eV and  $\epsilon_{s0} = 3.2$ eV to obtain the band gap of 3.2 eV at the $\Gamma$ $(k=0)$ point representing the band-edge in ZnO. We take the value of the $\langle s 0| \hat{p} | p 0 \rangle = 0.37$ au that match the  value in the realistic ZnO at the $\Gamma$ point. We use these parameters to solve \eq{eigval} to obtain band energies and  momentum matrix elements between the two bands. We do this for 500 $k$-points that sample the BZ and are obtained using the Born Von Karman boundary condition such that $k = \frac{2\pi j }{ 500 a}$, where $a=5.2$ \r{A} and $j$ is an integer $\in [-250,250)$. \Fig{kpbandstructure}a shows the band structure of the two-band model and \fig{kpbandstructure}b shows the field-free absorption spectrum with gray vertical line at 3.2 eV indicating the field free band-edge.

\section{Optical response of the laser-dressed two-band model}

\begin{figure}[htbp!]
    \centering
    \includegraphics[width=0.98\textwidth]{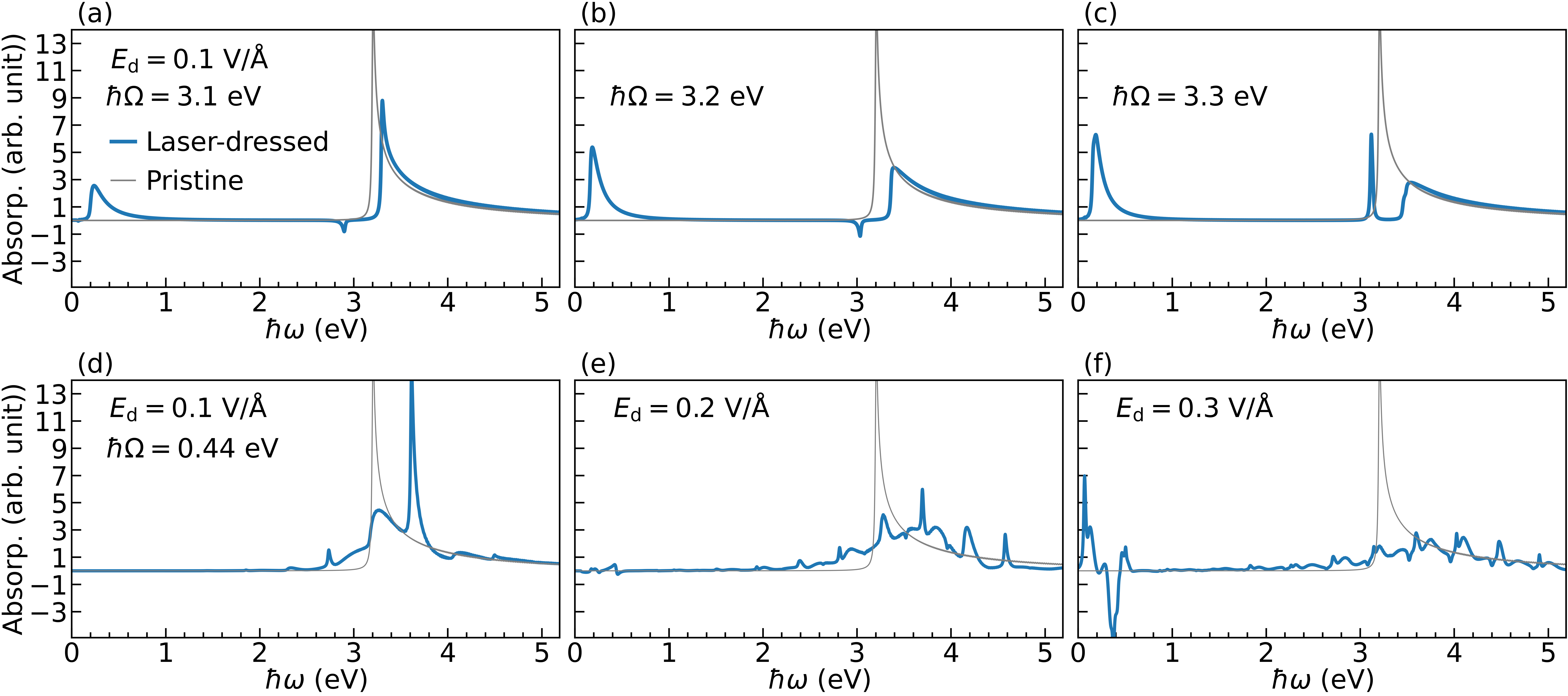}
    \caption{Optical absorption spectrum of the two-band model dressed with near-resonant (a)-(c) and off-resonance (d)-(f) light  of photon energy $\hbar\Omega$ and amplitude $E_{\mathrm{d}}$ (blue lines). The spectra  of the pristine one-dimension solid is shown in gray. }
    \label{twobandabsspectrum}
\end{figure}

Once the band energies and the momentum matrix elements are obtained for the two-band model, we compute the absorption spectrum using the theory of optical absorption of laser-dressed solids based on the velocity gauge light-matter interaction \cite{Tiwari2023}. We compute the spectra for drive laser parameters taken to be same as the ones for the ZnO computations in the main text. However, for the model the BZ is sampled using one-dimension grid of 500 $k$-points. In the spectra, each transition line is broadened using a Lorentzian function of width 0.02 eV and transition peaks before 0.06 eV are removed to account for finite discretization of the BZ.

\Fig{twobandabsspectrum} shows that the optical absorption spectra of the two-band model (blue lines) that is dressed with laser of photon energy $\hbar\Omega$ and amplitude $E_{\mathrm{d}}$. The spectra for the two-band model driven by band-edge near-resonant laser shown in \fig{twobandabsspectrum}a-c shows intense absorption features in the low-frequency. Also, for the off-resonance  case (shown in \fig{twobandabsspectrum}d-f), the spectrum for $E_{\mathrm{d}}=0.2$ V/\r{A} shows visible absorption and stimulated emission features in the low-frequency region whereas the spectrum for $E_{\mathrm{d}}=0.3$ V/\r{A} shows intense absorption and stimulated emission features in the low-frequency region. \Fig{twobandabsspectrum} shows that the two-band model qualitatively captures the same effect of strong driving as the bulk ZnO computations in the main text. However, the spectra for the two-band model show clear absorption sideband signals both below and above the band-edge (\fig{twobandabsspectrum}e-f) that are not present in the ZnO computations. These sidebands are visible in the minimalist model because of its lower dimensionality compared to bulk where spectral congestion obscures such sidebands. This shows that few-band models are not representative of experimentally relevant systems and computations with atomistically detailed model of full dimensionality are needed.

\section{Robustness of the low-frequency features in the presence of many-body interactions}

Below, we demonstrate that the low-frequency features remain robust even in presence of many-body interactions such as electron-electron and electron-nuclei. These interactions will naturally change the position and magnitude of these low-energy transitions as they modulate the many-body eigenstates of the system, but do not prevent the hybridization and the emergence of the low-frequency features in the optical absorption spectra of laser-dressed matter. 

To show this, we first develop a theory for the optical absorption of laser-dressed matter considering many-body matter Hamiltonian following the methodology in companion paper Ref. \cite{Tiwari2025}. We evaluate the resulting formula for the absorption coefficient in this case using electronic structure of matter obtained from numerically exact diagonalization of matter Hamiltonian. We then perform computations of the optical response of laser-dressed matter with varying strength of many-body interactions  in the matter Hamiltonian and show in each case the robustness of the low-frequency features.

\subsection{Theory of optical absorption of laser-dressed matter in a many-body electron-nuclear basis}

We first adapt our theory for the optical response of laser-dressed matter to a many-body matter Hamiltonian. That is, we do not employ any effective single-particle Hamiltonian like the ones developed in the manuscript from DFT. Instead, we suppose that we have access to the full many-body states of electrons and nuclei $\{\ket{\Psi_a}\}$, their energies $\{E_{a}\}$ and the transition dipoles $\{\bra{\Psi_b}\hat{\mu} \ket{\Psi_a}\}$. In this many-body basis, the Hamiltonian of the laser-dressed system reads $\hat{H}_\text{LD} = \sum_a E_a \ket{\Psi_a}\bra{\Psi_a} + \sum_{ab} \bra{\Psi_b}\hat{\mu}\ket{\Psi_a} \cdot \vect{E}_{\mathrm{d}}(t) \ket{\Psi_a}\bra{\Psi_b}$, where $\vect{E}_{\mathrm{d}}(t)$ is the electric field of the driving laser.
    
Following the derivation detailed in the companion paper Ref. \cite{Tiwari2025} for the optical absorption of laser-dressed systems, but employing the many-body states instead of the Bloch states, yields the following formula for the absorption coefficient at frequency $\omega$:
\begin{equation}
    \label{manybodyabs}
        A(\omega) = \frac{\omega }{4 \pi V \epsilon_{0} c n_{r}} \sum_{\alpha, \beta} \sum_{n} P _{\alpha} | \mu_{\alpha \beta}^{(n)}| ^ 2  [\delta(\mc{E}_{\alpha \beta} + n\hbar \Omega - \hbar \omega) - \delta(\mc{E}_{\alpha \beta} + n\hbar \Omega + \hbar \omega) ],
\end{equation}
where $V$ is the volume of the system, $\epsilon_{0}$ is vacuum's electric permittivity, $c$ the speed of light and $n_{r}$ the material's refractive index. The quantity $ \mc{E}_{\alpha\beta } = \mc{E}_{\alpha }-\mc{E}_{\beta } $ is the difference in quasienergy between Floquet mode $|\Phi_{\alpha} \rangle$ and $|\Phi_{\beta}\rangle$. In turn,
\begin{align}
    \mu_{\alpha \beta}^{(n)} & = \frac{1}{T}\int_{0}^{T} dt \bra{\Phi_{\alpha}(t)}  \hat{\mu}  \ket{\Phi_{\beta}(t)} e^{in\Omega t}
\end{align}
is the $n$-th Fourier component of the dipole matrix element between the \emph{many-body} Floquet modes. The population factors  
\begin{align}
  P _{\alpha}  & = \sum_i w_i |\langle \Psi_{i} | \Phi_{\alpha}(t_{0})\rangle | ^2 
\end{align} 
modulates the contribution from $\alpha$ Floquet mode to the optical response when the system is initially prepared in a density matrix  $\rho = \sum_{i} w_{i} |\Psi_{i} \rangle \langle \Psi_{i} |$. For thermal initial states, the coefficients $w_{i}$ would be the temperature-dependent Boltzmann probabilities. As before, the population factors depend on the initial state and the drive. 

As seen, the formal expression is closely related to the one in Eq. 1 in the manuscript with the key difference that the quasienergies $\mc{E}_{\alpha}$, population factor $P _{\alpha}$ and Fourier coefficients of the transition dipoles $\mu_{\alpha \beta}^{(n)}$ are those of the many-body Floquet states, instead of the single-particle states.

To evaluate Eq. \eqref{manybodyabs} in realistic systems is extraordinarily challenging as it is  simply not possible with present-day and even foreseeable computational resources to compute the many-body states of electrons and nuclei for realistic molecules and solids. However, the effect of many-body interactions on the low-frequency features can still be investigated by using a non-trivial model as shown below.

\subsection{Many-body Hubbard-Holstein model}
We consider a Hubbard-Holstein model, which is a workhorse of many-body physics. In this model, the regular tight-binding Hamiltonian is augmented with electron-electron and electron-nuclear interactions through Hubbard and Holstein terms respectively \cite{Schler2018, johnston_systematic_2010,esslinger_fermi-hubbard_2010}. The interplay between the two effects can generate rich phase diagrams with insulating, metallic, and superconducting phases \cite{fehske_quantum_2004,johnston_systematic_2010}.

Specifically, we use the Hamiltonian:
\begin{multline}
\label{eq:hamhh}
\hat{H}_\text{LD}=\sum_{n, \Delta} h_{nn} \hat{c}^{\dagger}_{n,\Delta}\hat{c}_{n,\Delta}+\sum_{\langle n,m\rangle, \Delta} h_{nm}(\hat{c}^{\dagger}_{n,\Delta}\hat{c}_{m,\Delta}+ \text{H.c.})+   U \sum_{n}\hat{c}_{n,+}^\dagger\hat{c}_{n,+}\hat{c}_{n,-}^\dagger\hat{c}_{n,-} \\ +  \alpha\sum_{n,\Delta}  \hat{X}_n \hat{c}^{\dagger}_{n,\Delta}\hat{c}_{n,\Delta}   + 
 \sum_i\left(\frac{\hat{p}_i^2}{2M_i} + \frac{1}{2} k_i  \hat{x}_i^2 \right) + \left(\sum_{n, \Delta}|e|n a \hat{c}^{\dagger}_{n,\Delta}\hat{c}_{n,\Delta} - \hat{\mu}_N \right)E_{\mathrm{d}}(t),
\end{multline}
where $c_{n\Delta}^\dagger$ ($c_{n\Delta}$) is a creation (annihilation) operator of electrons with spin $\Delta=\pm$ at a given site $n$ with $\langle n,m\rangle$ denoting nearest-neighbors, and $a$ the lattice constant. The first two terms on the right hand side of \eq{eq:hamhh} correspond to the kinetic energy of the electrons, while the on-site Coulomb repulsion between electrons is included in the third. The fourth term describes the Holstein electron-phonon interactions  to the phonon bath in the fifth term. Here $\hat{X}_n=\sum_i a_{in} \hat{x}_i$ is a collective phonon coordinate, $\{\hat{x}_i\}$ the phonon displacements from equilibrium position, $\{\hat{p}_i\}$ their conjugate momenta, $\{M_i\}$  the masses and $\{k_i\}$ their stiffness. The last term describes the light-matter interaction of the chain with the driving laser  $E_{\mathrm{d}}(t)$ in dipole approximation with $e$ denoting the electron charge, and $\hat{\mu}_N$ the nuclear dipole operator.

For definitiveness, in our computations we use a 
half-filled (neutral) open-boundary one-dimensional chain with $N=4$ sites, and the same number of fermions, coupled to one phonon. 
The model parameters are chosen to yield two bands as $N$ increases. Specifically, 
the lattice constant is taken to be $a=5$ \AA , the site energies $h_{nn}=2$ eV for odd sites and $0$ for even ones, and $h_{nm}=1$ eV between all nearest-neighbor sites. The nuclear parameters are as follows: mass $M=10,000$ (fs/\AA)\textsuperscript{2}, spring constant $k=60$ eV/\AA\textsuperscript{2}. This yields a frequency of $\sim12$ THz ($\sim 0.05$ eV which is in the typical range of optical phonon frequencies in solids.)
We take the phonon to represent the displacement of the 1st and 3rd 
sites from their periodic position, such that $\hat{X}_1=\hat{X}_3 = \hat{x}$ and $\hat{X}_2=\hat{X}_4 = 0$. In this model, the nuclear dipole operator becomes $\hat{\mu}_N = |e|(\sum_n na + 2\hat{x})$. This system represents a highly non-trivial  many-body problem that, importantly, remains computationally tractable to exact diagonalization techniques. Increasing the number of electrons or phonons beyond this limit makes the diagonalizations required for our Floquet analysis intractable due to the exponential growth of Hilbert space with the number of electronic, phononic and photonic degrees of freedom.

We elucidate the effects of the electron-electron correlation by varying the onsite Coulomb repulsion $U$ in the range of $0$ to $1$ eV, and the electron-phonon coupling  $\alpha$ in the range of $0$ to $2$ eV. The model captures electronic correlations and the strong vibronic interactions onset by photoexcitation of materials.

\subsection{Obtaining the many-body electron-nuclear states by direct diagonalization} 

To compute the many-body states, as developed in Ref. \cite{Hu2018}, we represent this Hamiltonian in matrix form by Jordan-Wigner transformation\cite{Jordan1928} of the creation and annihilation operators of the electrons and by discrete variable representation\cite{Light2007, Colbert1992} (DVR) of the nuclei. We obtain the many-body states for Hubbard-Holstein chains by directly diagonalizing this Hamiltonian. An advantage of this method is that it does not invoke any physical approximations, facilitating the interpretation of the results. Further, because it solves the many-body problem exactly, it can access regimes that are challenging for other methods. This includes materials containing light elements, where significant nuclear quantum effects invalidate the adiabatic approximation and the very concept of a (static) band structure, as well as strongly correlated systems where perturbative expansions to electron-electron interactions are inadequate.

The matrix representation of the Hubbard-Holstein Hamiltonian is constructed by tensor product of its electronic and nuclear components. To represent the fermionic creation and annihilation operators in matrix form, we adopt the Jordan-Wigner Transformation \cite{Jordan1928, Light2007, Hu2018}. In this transformation, the fermionic annihilation operators at site $n$ with spin $\Delta$ can be represented by a string matrix $e^{i\phi_{n, \Delta}}$ times the corresponding spin-$\frac{1}{2}$ Pauli lowering matrix $\sigma_{n, \Delta}$ at the same site and with the same spin, \textit{i.e.} $\hat{c}_{n,\Delta} \doteq e^{i\phi_{n, \Delta}}\sigma_{n, \Delta}$. Similarly, the fermionic creation operator is represented as $\hat{c}_{n,\Delta}^{\dag} \doteq \sigma_{n, \Delta}^{\dag}e^{-i\phi_{n, \Delta}}$, where $\sigma_{n, \Delta}^{\dag}$ is the spin-$\frac{1}{2}$ Pauli raising matrix at site $n$ with spin $\Delta$. Here, the phase matrix $\phi_{n, \Delta}$ contains the sum over all the occupation matrices to the left of ($n, \Delta$), \textit{i.e.} $\phi_{n, \Delta} = \pi \sum_{(k, \Delta') < (n, \Delta)}\sigma_{k, \Delta'}^{\dag}\sigma_{k, \Delta'}$. In our notation, we assume that the spin up $(+)$ is at the left of spin down $(-)$ for each site.

The spin number operator $\sigma_{n, \Delta}^{\dag}\sigma_{n, \Delta}$ is idempotent. Further, $\sigma_{n, \Delta}^{\dag}\sigma_{n, \Delta}$ with different ($n, \Delta$) commute. Thus, by Taylor expanding each component of $e^{i\phi_{n, \Delta}}=\prod_{(k, \Delta') < (n, \Delta)}e^{i\pi\sigma_{k, \Delta'}^{\dag}\sigma_{k, \Delta'}}$, the fermionic annihilation operators can be expressed as:
\begin{widetext}
\be
\label{eq:giant_matrix}
\begin{array}{ccccccccccccc} 
\hat{c}_{1+} & \doteq & \sigma_{1+} & \bigotimes & I_{1-} & \bigotimes& {I}_{2+} &\bigotimes & \cdots & \bigotimes & {I}_{N+} & \bigotimes & {I}_{N-} \\
\hat{c}_{1-} & \doteq & 1-2\sigma_{1+}^{\dag}\sigma_{1+} & \bigotimes & \sigma_{1-} & \bigotimes& {I}_{2+} &\bigotimes & \cdots & \bigotimes & {I}_{N+} & \bigotimes & {I}_{N-} \\
\hat{c}_{2+} & \doteq & 1-2\sigma_{1+}^{\dag}\sigma_{1+} & \bigotimes & 1-2\sigma_{1-}^{\dag}\sigma_{1-} & \bigotimes& \sigma_{2+} &\bigotimes & \cdots & \bigotimes & {I}_{N+} & \bigotimes & {I}_{N-} \\
\vdots\\
\hat{c}_{N-} & \doteq & 1-2\sigma_{1+}^{\dag}\sigma_{1+} & \bigotimes & 1-2\sigma_{1-}^{\dag}\sigma_{1-} & \bigotimes& 1-2\sigma_{2+}^{\dag}\sigma_{2+} &\bigotimes & \cdots & \bigotimes & 1-2\sigma_{N+}^{\dag}\sigma_{N+} & \bigotimes & \sigma_{N-} 
\end{array},
\ee
\end{widetext}
where $I_i$ is the $2\times2$ identity matrix in the $i$th subspace. The fermionic creation operators at site $n$ with spin $\Delta$ can be represented simply by replacing $\sigma_{n, \Delta}$ by $\sigma_{n, \Delta}^{\dag}$  in \eq{eq:giant_matrix}. In this way, the resulting operators satisfy the desired fermionic anti-commutation rules ($\{\hat{c}_{n,\Delta}, \hat{c}_{n',\Delta'}^{\dag}\} = \delta_{nn'}\delta_{\Delta\Delta'}, \{\hat{c}_{n,\Delta}, \hat{c}_{n',\Delta'}\} = \{\hat{c}_{n,\Delta}^{\dag}, \hat{c}_{n',\Delta'}^{\dag}\} = 0$) because spin-$\frac{1}{2}$ Pauli matrices satisfy the following relations: $\{\sigma_{n,\Delta}, \sigma_{n,\Delta}^{\dag}\} = 1$, $\{1-2\sigma_{n,\Delta}^{\dag}\sigma_{n,\Delta}, \sigma_{n,\Delta}^{\dag}\} = 0$ and $\{1-2\sigma_{n,\Delta}^{\dag}\sigma_{n,\Delta}, \sigma_{n,\Delta}\} = 0$.

The Fock space (in which the creation and annihilation operators are defined) includes all possible electronic number states. To reduce the computational effort, we project the electronic Fock space to a Hilbert space with a fixed $n_{e}$ number of electrons. This is possible since the Holstein-Hubbard Hamiltonian commutes with the electron number operator $\hat{n}_{e} = \sum_{n\Delta}\hat{c}_{n,\Delta}^{\dag}\hat{c}_{n,\Delta}$,
and thus the dynamics preserves $n_{e}$. The size of the net electronic basis is $\binom{2n_e}{n_e}$ for half-filled chains, which for $n_e=4$ evaluates to $70$ states. We further restrict the electronic space to $36$ states with zero projection of total spin along the quantization axis, $\hat{S}_z=0$, since the light-induced dynamics preserves the total spin. 

To represent the nuclear operators in terms of matrices, we employ the Discrete Variable Representative (DVR) as proposed in Ref. \cite{Colbert1992}. DVR methods have been proved to be highly accurate to solve a variety of problems in molecular quantum dynamics and vibration-rotation spectroscopy\citep{Light2007}.  For simplicity, we consider one nuclear degree of freedom. For this degree of freedom, a basis consisting of equally spaced grid points, $\{\ket{i}\}$, at positions $x_i$ is employed.  The matrix elements of the kinetic energy operator in this basis are:
\begin{equation}
\label{K_DVR}
\bra{i}\hat{T}\ket{i'} = \frac{\hbar^2(-1)^{i-i'}}{2M(\Delta x)^2}\left\{
       \begin{array}{cccccc}
       \pi^2/3&, & \quad   &i&=&i' \\
       \frac{2}{(i-i')^2}&,& \quad &i&\neq& i'
       \end{array}
\right\},
\end{equation}
where $\Delta x$ is the grid spacing (chosen to be $0.005$\AA) and the grid spans the range of positions from $-0.08$ to $0.08$ \AA. This range corresponds to about four times the typical displacement of the ground vibrational state, $\Delta x=\sqrt{\frac{\hbar}{2M\omega}}$, where $\omega = \sqrt{K/M}$ is the phonon frequency. Correspondingly, in DVR the matrix elements of the position dependent functions $V(\hat{x})$ are:
\begin{equation}
\label{V_DVR}
\bra{i}V(\hat{x})\ket{i'} = V(x_i)\delta_{ii'}.
\end{equation}
For our problem, the total dimension of vibronic Hilbert space is therefore $36\times33=1188$. All results presented here have been tested for convergence in the grid spacing and the range of space considered in the simulation.

Once the Hamiltonian is constructed in this fashion, we directly diagonalize it to obtain the electron-nuclear Hamiltonian eigenstates for this problem, and to compute the matrix elements of the dipole operator. The eigenenergies, and the dipole matrix elements are then used as input in the non-equilibrium optical absorption evaluated using Eq. \eqref{manybodyabs}.

\subsection{Computation of the non-equilibrium absorption spectrum}

As shown in Figure 3 of the main text, both resonant and non-resonant driving of the system leads to the formation of Floquet replicas that can energetically overlap. This induces hybridization of the Floquet states and formation of hybrid Floquet states that display avoided crossings as a function of laser parameters. Probing this laser-dressed band structure leads to the low-frequency features, that in turn, provide purely optical signatures of Floquet states. In nature, the optical response of laser-dressed materials arises due to many-body Floquet replicas and the hybridization of the many-body Floquet states of electrons and nuclei. Therefore, electron-electron and electron-nuclei interactions will naturally change the position and magnitude of these low-energy transitions as they modulate the many-body eigenstates of the system, but do not prevent this hybridization and the emergence of the low-frequency features in the optical absorption spectra of laser-dressed matter from occurring. 

\begin{figure}[h!]
    \centering
    \includegraphics[width=0.99\textwidth]{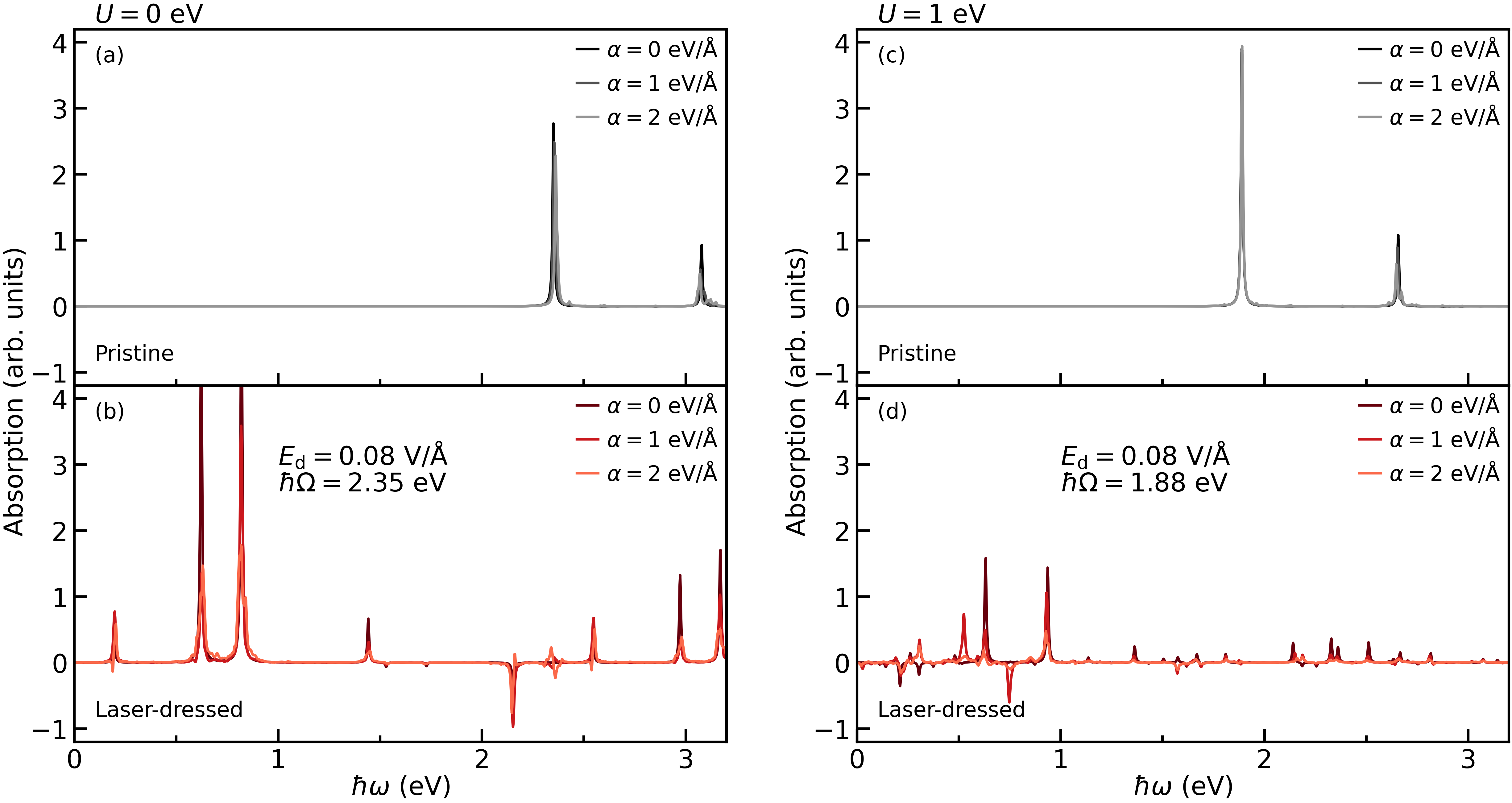}
    \caption{Influence of increasing the electron-phonon couplings $\alpha$ on the equilibrium (top panels) and non-equilibrium (bottom panels) optical absorption spectra of the Hubbard-Holstein model for (a)-(b) $U=0$  and (c)-(d) $U=1$ eV. Note that the low-frequency features due to Floquet state hybridization remain robust to electron-phonon couplings.}
    \label{varyalpha}
\end{figure}

\begin{figure}[h!]
    \centering
    \includegraphics[width=0.99\textwidth]{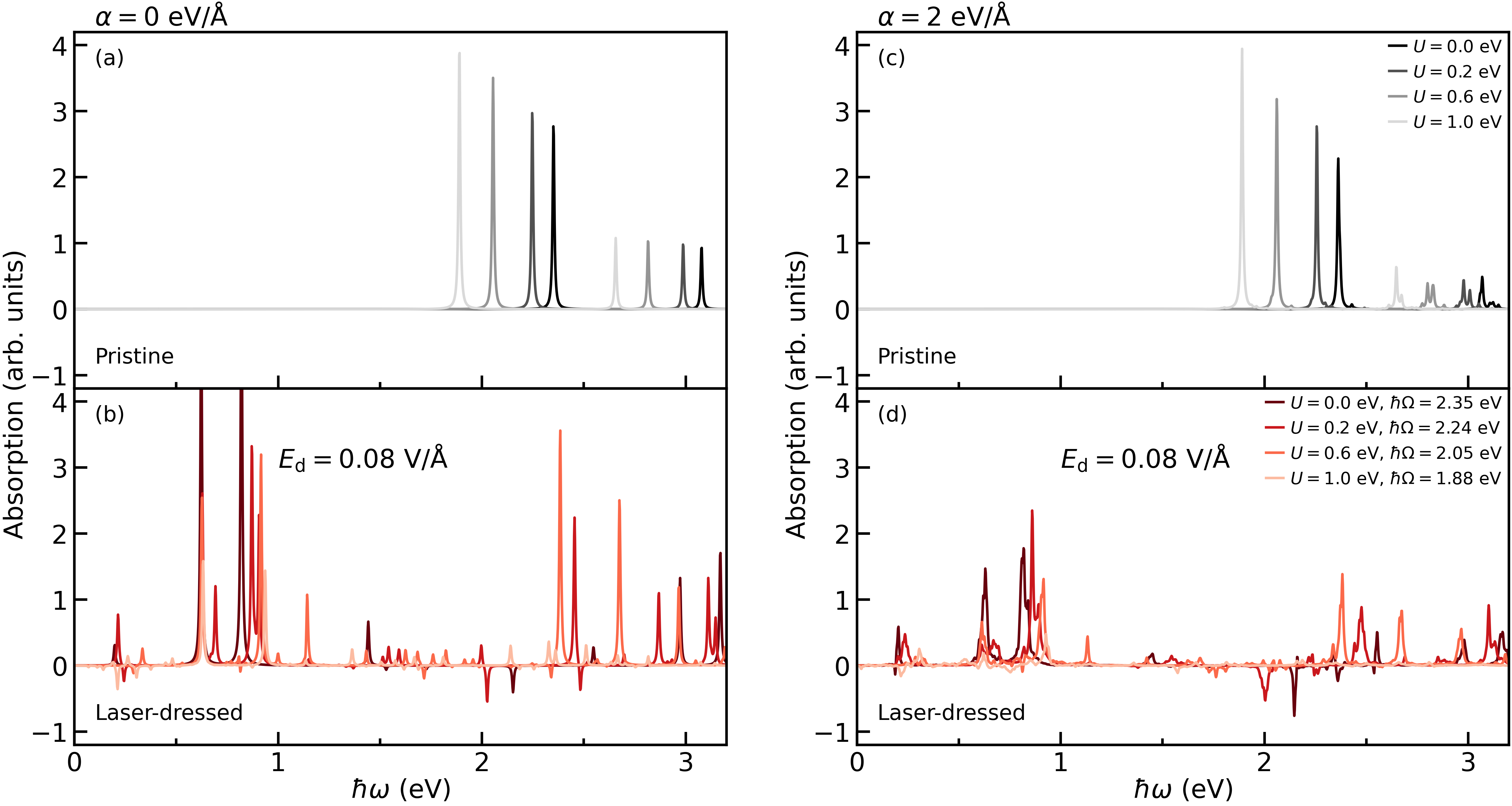}
    \caption{ Influence of increasing the Hubbard repulsion $U$ on the equilibrium (top panels) and non-equilibrium (bottom panels) optical absorption spectra  of the Hubbard-Holstein model for (a)-(b) $\alpha=0$  and (c)-(d) $\alpha=2$ eV/\AA. Note that the low-frequency features due to Floquet state hybridization remain robust to many-body electronic repulsion.}
        \label{varyu}
\end{figure}

Below we show for a resonantly driven chain how the low-frequency features are changing due to varying each of the following: the electron-phonon coupling $\alpha$ (\fig{varyalpha}), the onsite Coulomb repulsion $U$ (\fig{varyu}), and the temperature  (\fig{varytemp}). Additionally in \fig{varyintensity}, we investigate the influence of explicit many-body correlations on the laser-controllability of  the low-frequency features (as shown in Fig. 3a of manuscript). Converged computations have been obtained using 33 Floquet Fourier basis and the transition lines are broadened using a Lorentzian broadening of width 0.01 eV.

\subsubsection{Influence of electron-phonon couplings.}
\Fig{varyalpha} shows the influence of increasing the electron-phonon couplings $\alpha$ on the equilibrium (top panels) and non-equilibrium (bottom panels) optical absorption spectra for Fig. \ref{varyalpha}(a)-(b) $U=0$  and Fig. \ref{varyalpha}(c)-(d) $U=1$ eV. For $U=0$, the spectrum of pristine matter shows sharp absorption features at $\hbar\omega=2.35$ eV and 3.1 eV. The same qualitative features are observed when electronic repulsion $U$ is turned on but with different transition frequencies and spectral intensities.  Upon laser-dressing, several new spectral features emerge. In particular, due to Floquet state hybridization, the  characteristic low-frequency absorption features emerge around $\hbar\omega\sim 0.2$ eV.

Under equilibrium and non-equilibrium conditions, these isolated peaks become a manifold of transitions once $\alpha\ne 0$. This is because the electron-phonon interactions turns the purely electronic problem into a electron-vibrational (vibronic) problem and the different sidebands correspond to different transitions among vibronic states, with the 0-0 transition remaining the most prominent. Increasing $\alpha$ makes the vibronic Franck-Condon progression increasingly more prominent. If these spectral features are not optically resolved, then they effectively lead to broadening of the absorption features like the one used in the main paper. Note that  the electron-phonon interactions do not prevent the hybridization between Floquet states that lead to the low-frequency transitions from happening. They just introduce additional vibronic levels that can hybridize and, naturally, change the details of the optical response. If these levels are not spectrally resolved, the net effect is to become a source of broadening to the non-equilibrium optical absorption spectra.

To conclude: the low-frequency transitions that can be used to detect Floquet states by purely optical means are robust to increasing electron-phonon interactions, in the presence and absence of Hubbard repulsion terms.  Electron-phonon interactions can be approximately modeled as a source of broadening to the electronic transitions.

\subsubsection{Influence of electronic repulsion.}
\Fig{varyu} shows the influence of increasing the electron repulsion $U$ on the equilibrium (top panel) and non-equilibrium (bottom panel) optical absorption spectra for (a)-(b) $\alpha=0$  and (c)-(d) $\alpha=2$ eV\AA. In pristine matter, increasing $U$ changes the position of the main spectral peaks and the intensity of the optical transitions in the presence and absence of electron-phonon interactions.  For the laser-dressing we choose a laser amplitude $E_{\mathrm{d}}=0.08$ V/\r{A}  at resonance with the lowest-energy transition of the pristine system in each case.

As before, upon laser-dressing, the hybridization of Floquet modes induce a prominent low-frequency feature at $\hbar\omega\sim 0.2$ eV. Here again, increasing $U$ modulates the magnitude and frequency of these low frequency transitions but does not prevent the hybridization from occurring. This behavior is robust in the presence of electron-phonon couplings as well.

To conclude: the low-frequency transitions that can be used to detect Floquet states by purely optical means are robust to increasing electron-electron interactions, in the presence and absence of electron-phonon interactions.  This is because they arise from the hybridization of many-body Floquet states and thus the hybridization is modulated, but not prevented, by electron correlations. 

\begin{figure}[h!]
    \centering
    \includegraphics[width=0.87\textwidth]{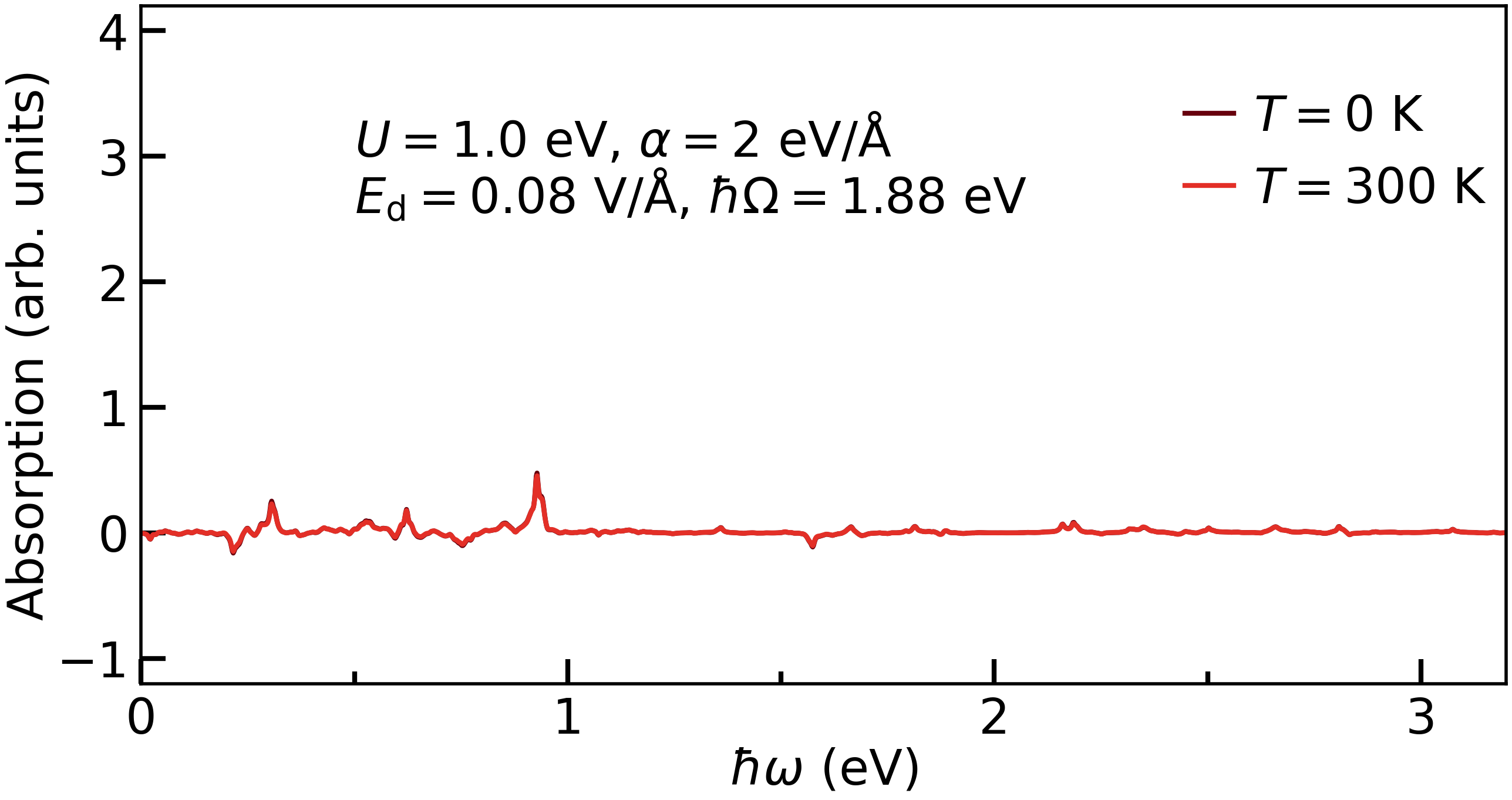}
    \caption{ Influence of  increasing temperature on the non-equilibrium optical absorption of the laser-dressed Hubbard-Holstein model }
        \label{varytemp}
\end{figure}

\subsubsection{Influence of temperature.} We characterize the influence of initial temperature on the non-equilibrium optical absorption spectra of the Hubbard-Holstein model for resonant driving. As shown in \fig{varytemp}, temperature plays a minor role in the non-equilibrium optical spectra. This is because the energy of a vibrational transition for this model is $\sim 0.05$ eV (which is a typical value for vibrations in molecules and optical phonons in materials). This is almost twice the thermal energy is 0.0259 eV at 300K. Therefore, at room temperature the system is mostly in the vibrational ground state. Increasing the temperature further will introduce additional peaks in the vibronic progression that will effectively increase the broadening of the electronic transitions due to limited spectral resolution. However, as demonstrated in \fig{varytemp}, the low-frequency features survive such spectral broadening.  

\begin{figure}[h!]
    \centering
    \includegraphics[width=0.87\textwidth]{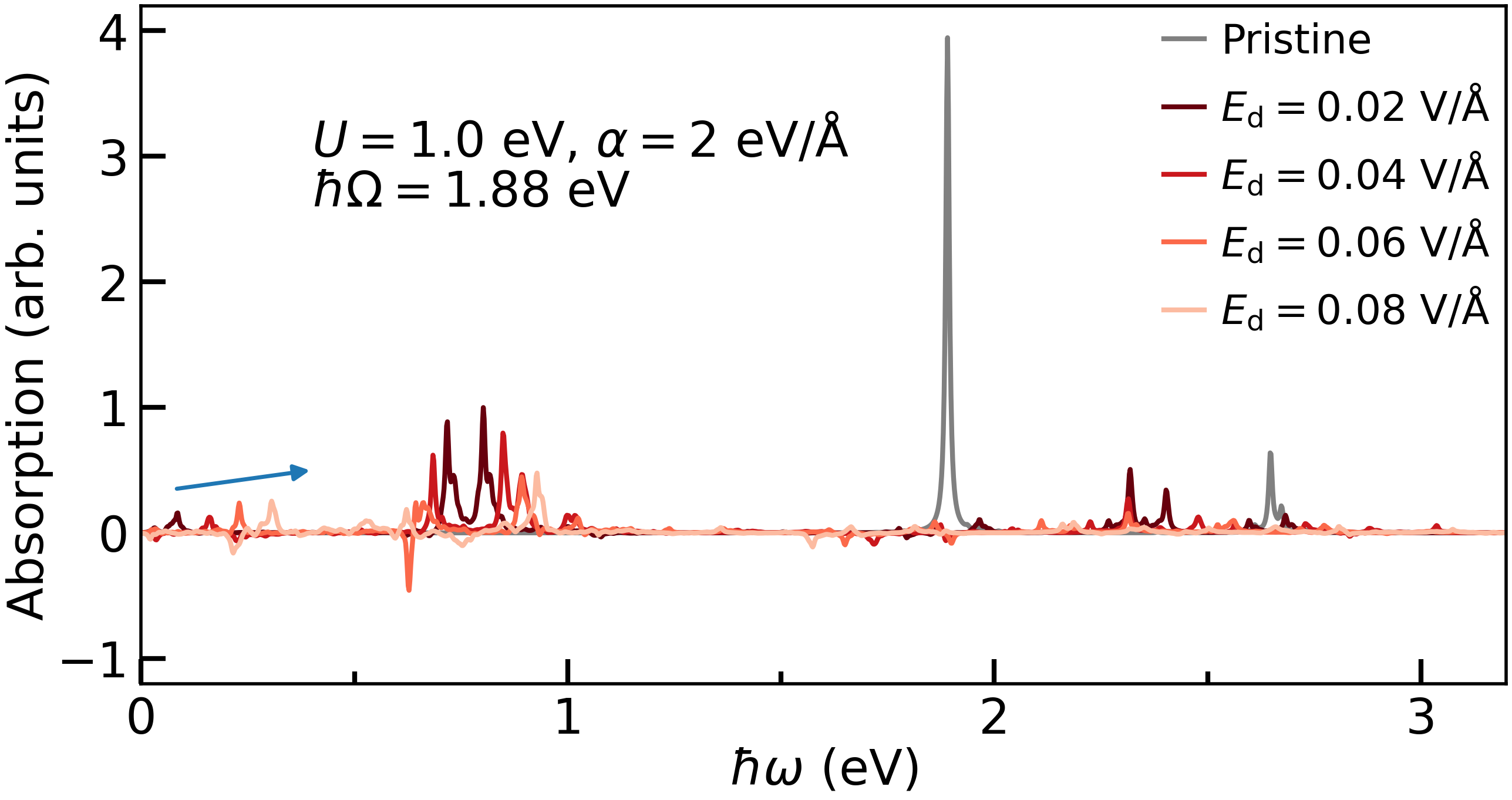}
    \caption{ Influence of driving-laser intensity on the low-frequency transitions that emerge in laser-dressed materials in the presence of Hubbard repulsion $U$ and electron-phonon coupling $\alpha$ for the Hubbard-Holstein model. Increasing the driving-laser intensity blue-shifts these transitions shown by blue arrow as it increases the gap between Floquet modes due to their hybridization. }
        \label{varyintensity}
\end{figure}

\subsubsection{Influence of driving-laser intensity.} Lastly, for resonant driving, we showed in Fig. 3c in the manuscript that these low-frequency transitions are laser-controllable and can be blue shifted by increasing the laser amplitude. This is because the laser intensity determines the strength of hybridization and, thus, the energy gap that emerges between the Floquet states due to such hybridization. As shown in \fig{varyintensity} (blue arrow), this feature is also clearly exhibited by the Hubbard-Holstein model even in the presence of the electron correlations and electron-phonon interactions. This demonstrates that the physical origin of these low frequency transition is identical to the one we isolated through first-principle computations in ZnO.

Summarizing, we clearly showed, through  computations in a non-trivial many-body model with explicit electron-electron  and electron-nuclear interactions, that the low-frequency features remain robust to many-body interactions. We further showed that these features arise due to the hybridization of  many-body Floquet states.

\section{Absorption spectrum of laser-dressed Z\lowercase{n}O with different spectral broadening}

\begin{figure}[htbp!]
    \centering
    \includegraphics[width=0.98\textwidth]{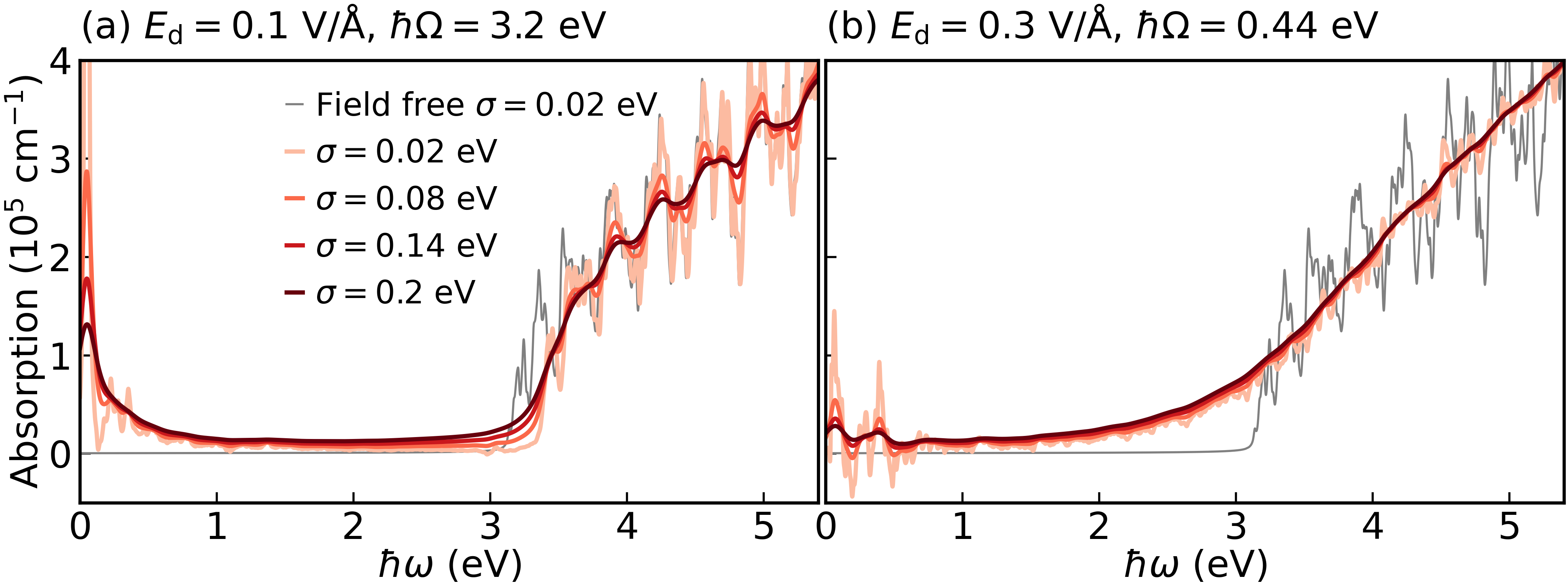}
    \caption{Optical absorption spectrum of the ZnO dressed with (a) near-resonant, and (b) off-resonance light  of photon energy $\hbar\Omega$ and amplitude $E_{\mathrm{d}}$. Lorenztian broadening of width $\sigma$ is used to broaden the spectral lines.}
    \label{sigmazno}
\end{figure}

Under experimental conditions, electron-nuclear interactions can introduce additional fine structure to the transition lines in the absorption spectrum of laser-dressed solid that, if it is not spectrally resolved, leads to spectral broadening of the participating transitions through Frank-Condon like progressions \cite{Szidarovszky2019}. We model the resulting spectral broadening using a Lorentzian function of a particular full-width at half maxima $\sigma$ for each transition line. \Fig{sigmazno} shows the optical absorption spectrum of laser-dressed ZnO for values of drive field amplitude $E_{\mathrm{d}}$ and photon energy $\hbar\Omega$ with prominent low-frequency features in Fig. 2 of main text. The different lines in red represent the spectra for various $\sigma$ and the grey line represents the field-free absorption spectra for $\sigma=0.02$ eV.  The low-frequency features in the range $\hbar\omega \in [0, 0.6]$ eV are clearly visible in \fig{sigmazno}a-b which correspond to resonant and off-resonance driving laser conditions respectively. As can be seen, as $\sigma$ increases, some of the fine structure in the low-frequency region visible for low broadening ($\sigma =0.02$ eV) smooths out as $\sigma$ is increased. However, the low-frequency signals remain prominent even for very high values of $\sigma=0.2$ eV which correspond to a ultrafast dephasing timescale of $\tau = \frac{\hbar}{\sigma}=3.25$ fs.

